# State Space Models Naturally Produce Time Cell and Oscillatory Behaviors and Scale to Abstract Cognitive Functions

Sen Lu[1,§], Xiaoyu Zhang[1,§], Mingtao Hu[1], Eric Yeu-Jer Lee[1], Soohyeon Kim[1], Wei D. Lu[1,*]

[1]Department of Electrical Engineering and Computer Science, University of Michigan, Ann Arbor, MI 48105, USA.

[§]These authors contributed equally.

*Correspondence to: wluee@umich.edu.

These authors contributed equally: *Sen Lu*, *Xiaoyu Zhang*
Corresponding Email: wluee@umich.edu

## ABSTRACT

A grand challenge in modern neuroscience is bridging the gap between the microscale neural circuits and cognitive functions. While extensive knowledge exists about neuronal connectivity and biophysics, how low-level phenomena give rise to abstract behaviors remains largely unresolved. Here, we propose that a mathematical framework based on State Space Models offers mathematical sufficiency and computational plausibility to generate response profiles that resemble time cells, ramping activity, and oscillatory responses. Our results suggest that such behaviors can emerge from a structured multi-timescale complex rotational basis, which enables optimized history compression and naturally generates structured temporal dynamics even prior to training, mirroring recent findings in biological circuits. Learning exploits and adapts to the pre-existing temporal basis for specific tasks. We show that the model can readily scale to abstract cognitive functions such as event counting, supporting its use as a computationally tractable framework for understanding neural activities.

## INTRODUCTION

The last few decades have witnessed extraordinary progress in neuroscience, particularly in mapping the brain's intricate wiring diagram. We can now trace pathways with synaptic resolution and record the activity of thousands of neurons simultaneously during complex behaviors[1,2]. Yet, questions still remain on how the collective activity of these meticulously mapped circuits gives rise to cognitive functions such as memory, perception, or the recognition of an abstract concept[3].

Among these cognitive functions, the ability to encode and distinguish temporal sequences is fundamental for behaviors, ranging from interval timing to episodic memory[4]. Within experimental neuroscience, a key discovery is the existence of time cells[5], neurons found in regions including the hippocampus, prefrontal cortex, and striatum[6]. These cells become active sequentially, each firing at a specific moment during a timed delay period[7]. Together, their activity effectively marks the flow of time. Alongside these, researchers have identified other properties, such as ramping cells, whose firing rates monotonically increase or decrease over a timed interval[8], and persistent neural oscillations, which are also thought to be critical for time-keeping[9]. These discoveries offer a powerful insight into the neural basis of temporal memory, but they also raise deeper questions: What are the underlying mechanisms of these temporal codes? What is the role of learning vs. innate structures given recent observations that these sequences can flexibly recalibrate, or remap, to adapt to different durations without re-training? A model that can provide theoretical insight into these questions remains a key objective in the field.

In the effort to model phenomena like time cells, prevailing approaches have often operated from two opposing perspectives. Approaches from the bottom-up start from the neuron. At the microscale, since detailed biophysical models like the Hodgkin–Huxley model (HH model)[10] provide an exquisite description of how action potentials are generated based on the dynamics of ion channels. These models are grounded in molecular processes but are computationally intensive and difficult to scale into large networks capable of cognitive tasks[11,12]. On the other hand, the top-down approach utilizes concepts of neural networks such as recurrent neural networks (RNNs)[13], convolutional neural networks (CNNs)[14] and transformer-based Large Language Models (LLMs)[15], to describe cognitive processes. Studies using artificial neural networks to support neuroscience findings have recently moved beyond phenomenological reproduction, demonstrating that time-cell-like sequences, ramps, and oscillations may naturally emerge in RNNs and Deep Reinforcement Learning (DRL) agents trained on timing and working memory tasks[16,17]. For example, Lin et al.[17] established that these dynamics are not merely by-products of training but actively support value estimation and policy computation, with time and ramping cells carrying decodable information about temporal intervals and stimulus identity. Similarly, work by Zhou et al.[16] has shown that trained RNNs can generate realistic spiking sequences akin to biological time cells. However, while these networks confirm that temporal codes can emerge from recurrent activity, they often lack an explanation for why biological circuits favor specific representations over others. Additionally, time cells exhibit complex adaptive behaviors beyond what are typically expressible through simple recurrent circuits. While some neural circuits rescale or stretch their temporal receptive fields[18,19], others exhibit remapping[7,20]. For instance, Taxidis et al.[21] demonstrated that hippocampal time-cell ensembles remap rather than stretch when delays are extended in working memory tasks.

Recent lines of evidence also challenge the assumption that temporally organized sequences are exclusively the product of learning from scratch[22], pointing instead to the critical role of preconfigured, intrinsic network architectures. While the Entorhinal Cortex (EC) is often believed to be the driver of downstream temporal codes, Heys and Dombeck[23] imply that CA1 and CA3 networks possess an inherent capacity to generate sequential firing dynamics even when EC input is compromised, although these intrinsic sequences display reduced stability. This suggests that the fundamental machinery for temporal organization resides within the local circuit dynamics themselves. Expanding on this view, several studies have found the emergence of preconfigured brain dynamics independent of sensory experience[24–26]. Recently, van der Molen et al.[27] observed similar intrinsic dynamics within human brain organoids, reporting the emergence of preconfigured neuronal firing sequences in the human brain, systems entirely devoid of sensory experience. Together, these findings suggest that temporal processing may not require de novo learning but rather exploits pre-existing, intrinsic dynamics that are biologically hardwired into the neural microcircuitry.

Here we propose that state space models (SSMs) with S5/HiPPO pre-configuration, originally developed in the field of control theory and recently used as a generative model for sequence processing[28,29], can help reconcile millisecond time-scale, microscale neuronal behaviors with the long time-scale, macroscale functions. In particular, we show that the network, through pre-configured structural basis, can naturally produce desired neural behaviors such as time cells and oscillations even without learning. During learning, the state-transition dynamics largely retain this pre-existing structure, whereas input modulations and readout projections adapt to the task. Hence, learning exploits and adapts to a substantial pre-existing temporal basis rather than constructing temporal codes de novo.

Unlike standard RNNs that operate in discrete steps, SSMs are rooted in continuous-time differential equations, making them conceptually similar to biophysical models of neuronal dynamics. In this view, a neuron's internal state—encompassing membrane potential and ion channel kinetics—is a continuous vector that evolves based on its history and external inputs. Formally, a linear time-invariant SSM is governed by (1), (2):

$$\mathbf{h}'(\mathrm{t}) = \mathbf{A}\mathbf{h}(\mathrm{t}) + \mathbf{B}\mathbf{x}(\mathrm{t}), \tag{1}$$

$$\mathbf{y}(\mathrm{t}) = \mathbf{C}\mathbf{h}(\mathrm{t}), \tag{2}$$

where, $\mathbf{h}(t) \in \mathbb{C}^N$ is the hidden state, $\mathbf{x}(t) \in \mathbb{R}^M$ is the input, $\mathbf{y}(t) \in \mathbb{R}^P$ is the output, $\mathbf{A}, \mathbf{B}$, and $\mathbf{C}$ are learnable parameters. Matrix $\mathbf{A}$ and $\mathbf{B}$ govern the state's evolution and its response to input.

Recent theoretical advances demonstrate that initializing $\mathbf{A}$ according to the HiPPO framework (High-order Polynomial Projection Operators) optimally projects continuous history onto basis functions of orthogonal Legendre polynomials[30]. Here we adopt a diagonalized version of this approach with complex-valued coefficients[31], denoted as $\mathbf{\Lambda}$. For computational implementation, this continuous system is discretized with a step size $\Delta t$, yielding the recurrence (3):

$$\mathbf{h}(t + \Delta t) = \overline{\mathbf{\Lambda}}(\Delta t)\boldsymbol{h}(t) + \overline{\mathbf{B}}\,\mathbf{x}(t + \Delta t), \tag{3}$$

where $\overline{\mathbf{\Lambda}}(\Delta t) = \exp(\mathbf{\Lambda}\,\Delta t)$ and $\overline{\mathbf{B}} = \mathbf{\Lambda}^{-1}(\exp(\mathbf{\Lambda}) - \mathbf{I})\mathbf{B}$. Due to the diagonal structure of $\overline{\mathbf{\Lambda}}$, individual state evolves independently following rotational dynamics before being mixed at the readout layer.

This mathematical formulation leads to our main hypothesis: these pre-configured oscillatory states could act as a structural basis for temporal cognition, and SSM offers a mathematical framework in which such dynamics are sufficient to serve as computational primitives.

As illustrated in Figure 1, this scheme interprets for time cells and other temporal codes as basis functions for input compression and shows how such codes could arise from rotational dynamics governed by equation (1). Incoming sensory streams, such as those from the EC, could be projected onto these basis functions within the hippocampal network, with the network's output then used for downstream tasks.

Importantly, we show that temporal receptive fields can be obtained from optimal initialization rather than training, i.e., an SSM network initialized according to the HiPPO framework[30] spontaneously exhibits time-cell sequences and ramping activity even before training. This result supports recent neurophysiological findings in human brain organoids, in which backbone firing sequences emerge intrinsically from a preconfigured 3D neural architecture, independent of sensory experience[27]. On the other hand, we found that randomly initialized SSMs do not produce such structured dynamics, even after extensive training, despite similar task performance. We suggest that optimal initialization can serve as a structural prior, a preconfigured architecture that allows such systems to rapidly capture long-range temporal dependencies at little learning cost.[32] From an evolutionary perspective, this framework suggests the computational plausibility that neural circuits approximate such basis functions, which would allow organisms to efficiently encode the complex temporal histories that behavior requires. Under this view, learning exploits and adapts these basis functions instead of constructing from scratch. While the input and readout adapt to map sensory inputs onto the basis functions and to select compressed history states for downstream tasks, the state-transition dynamics that define the basis functions retain much of their initial structure.

We then extend this mechanism to abstract cognitive functions, showing that continuous state-space trajectories can naturally transform repeating sensory inputs into decodable states, such as the event identity of repetition. In this work, we show that these findings collectively suggest that optimal state-space dynamics provide a computationally tractable and scalable framework for understanding of temporal cognition.

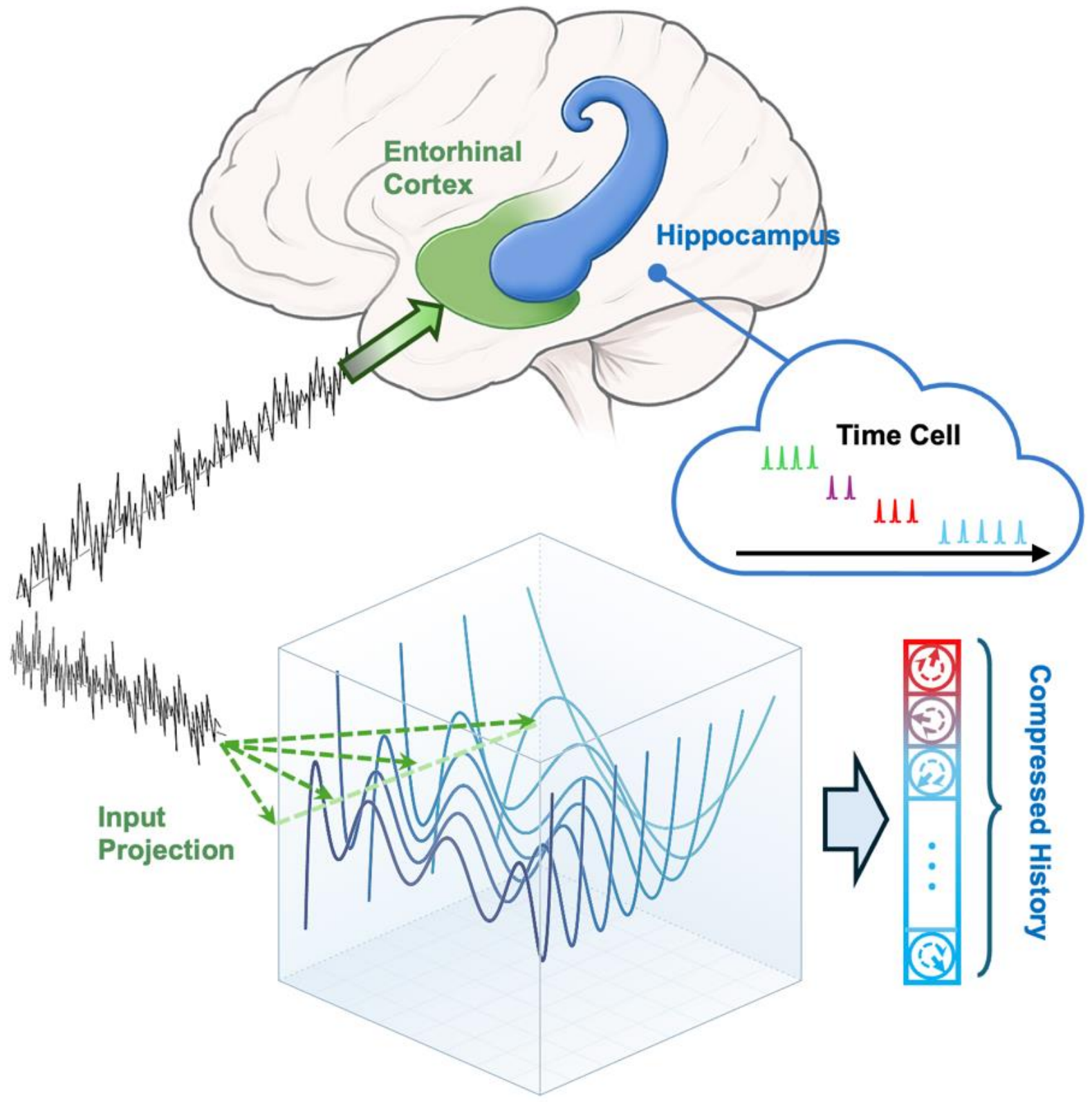

Figure 1 | A state-space perspective on neural temporal sequences. Top: sensory inputs (*e.g.*, from the Entorhinal Cortex) are processed by intrinsic, pre-configured circuit dynamics manifested as time cells that naturally compress input history. Synaptic plasticity then adapt this basis by sampling fromspecific temporal codes. Bottom: Mathematical implementation via an SSM. The HiPPO initialization serves as the mathematical proxy for the pre-configured structures, projecting continuous inputs onto a set of optimal basis functions (visualized as trajectories within the cube). The temporal information is compressed in a state vector, from which the network learns to select specific features for downstream tasks.

# RESULTS

## Three-Stimuli Temporal Discrimination Task

To test our hypothesis that pre-configured basis functions naturally lead to time cell behaviors, we trained an S5-based[31] agent on a temporal discrimination task using actor-critic deep reinforcement learning. As depicted in Fig. 2a, the agent's goal is to correctly identify the longest of three stimuli presented sequentially. The agent is trained through RL, similar to Lin et al.[17], who utilized a Long Short-Term Memory (LSTM) module and trained the agents using the actor-critic algorithm[33] to identify the longer sequence of two stimuli. Compared with prior studies that used the standard two-stimuli task, we chose a more challenging three-stimuli configuration. This shift was motivated by preliminary tests indicating that the two-stimuli task presented an insufficient challenge for both LSTM and SSM architectures. Both networks rapidly reached ceiling performance, making it difficult to effectively distinguish their capabilities. The increased complexity of the three-stimuli task was therefore necessary to provide a richer regime for observing the emergence of structured temporal dynamics. In the three-stimuli case in our study, in addition to the final reward for the correct choice, the learning process is guided by rewarding the agent for choosing the longer of the first two stimuli (Supplementary Methods 1.1). This intermediate reward aligns with the discovery that midbrain dopamine influences the perception of time in neurons[34]. The SSM layer output is then processed by the value network, which predicts the expected reward at the current state, and a policy network that predicts the action needed to be taken. Notably, the SSM layer communicates with the downstream policy and value networks through spiking signals, mimicking real neuron communication. A spike is generated from a hidden state unit when the real part of its activation value crosses a set threshold, analogous to a biological neuron's firing mechanism. Since the spiking function is non-differentiable, we employ surrogate gradient to enable backpropagation. We selected a Sigmoid-shaped surrogate function because its smooth, bell-shaped profile provides a continuous gradient that balances the trade-off between gradient sparseness and training stability. Empirical evaluations during our hyperparameter search indicated that this smoother transition led to faster convergence and higher task accuracy compared to piecewise linear alternatives. After training, the agent can correctly identify the longest sequence with an average accuracy of $88.1 \pm 1.9\%$ (10 independently trained agents). The LSTM comparison is provided in Supplementary Note 3, Supplementary Fig. 4 and Supplementary Table 1.

We visualize the typical network's internal dynamics by showing its states. Fig. 2b displays heatmaps of the real part of the state activities during the delays, averaged across 1000 trial episodes, during the two delay periods between the 3 stimuli. A clear sequential cascade emerges when these state units are sorted by the time of their peak activation, following Lin et al.[17]. Different units become selectively active at different moments, effectively cascading the entire temporal interval, replicating the time cell sequences observed in brain regions like the hippocampus and prefrontal cortex[7]. The heatmap also reveals high-frequency patterns in the remaining population (bottom rows). These higher-frequency oscillations, together with time cells and ramping cells, share the same mathematical form and could arise from the same underlying dynamics, but would manifest differently in experiments because their oscillations span different timescales. Together, they form a basis capable of efficiently approximating signal history over a wide range of delays[35]. Since the internal state dynamics (specifically, matrix $\mathbf{\Lambda}$ ) remain nearly unchanged during training, these high-frequency modes are preserved as basis functions, although we focus here on the lower-frequency components that manifest as explicit time cells. The lesion study (Supplementary Note 1 and Supplementary Figs. 1 and 2) also shows that high-frequency units are less important for this temporal discrimination task. The sorting of the state space into low-frequency time and high-frequency cells applies to all tasks presented and the details can be found in Method section.

Furthermore, this integrated architecture spontaneously produced a diverse set of temporal-coding mechanisms. Across the two delay periods, we found that 21.0% of the population emerged as time cells, while 61.8% exhibited monotonically increasing or decreasing activity (ramping cells) across 5000 trial episodes, as illustrated by the example tuning curves in Fig. 2c. All unsorted tuning curves are shown in Supplementary Note 2 and Supplementary Fig. 3 and the specific criteria used to classify these distinct functional cell types are also detailed in the Methods section.

In essence, since the S5 network uses a diagonal state matrix $\mathbf{\Lambda}$, the state update during the delay $\mathbf{h}(t+1) = \overline{\mathbf{\Lambda}}\mathbf{h}(t)$ simplifies to a set of independent equations for each state element: $h_i(t+1) = \overline{\lambda_i} h_i(t)$. Here, $\lambda_i$s are the diagonal elements of the discretized state matrix $\mathbf{\Lambda}$, which are learnable complex parameters. During the delays, where information must be preserved without decay, the network learns values of $\overline{\lambda_i}$ with a modulus of 1. Mathematically, this operation is a pure rotation in the complex plane and effectively performs a spectral decomposition of the network dynamics, treating them as independent latent oscillatory states[36]. This perspective aligns with the neural manifold hypothesis[37], which suggests that the collective activity of neural populations is governed by low-dimensional latent dynamics, and that individual neurons act as specific projections of these latent states[38]. Fig. 2d illustrates this mechanism in which the diverse temporal profiles are simply projections of these underlying rotations onto the real axis. Within this model, a time cell and a ramping cell may not represent distinct biological entities, but rather distinct manifestations of the different phases of rotational process projected onto axes. A state rotating at a specific frequency will appear as a time cell if its trajectory crosses the positive real axis during the delay. However, if it rotates from the positive real axis to the negative real axis or in the opposite direction, it reproduces the ramping cell behavior.

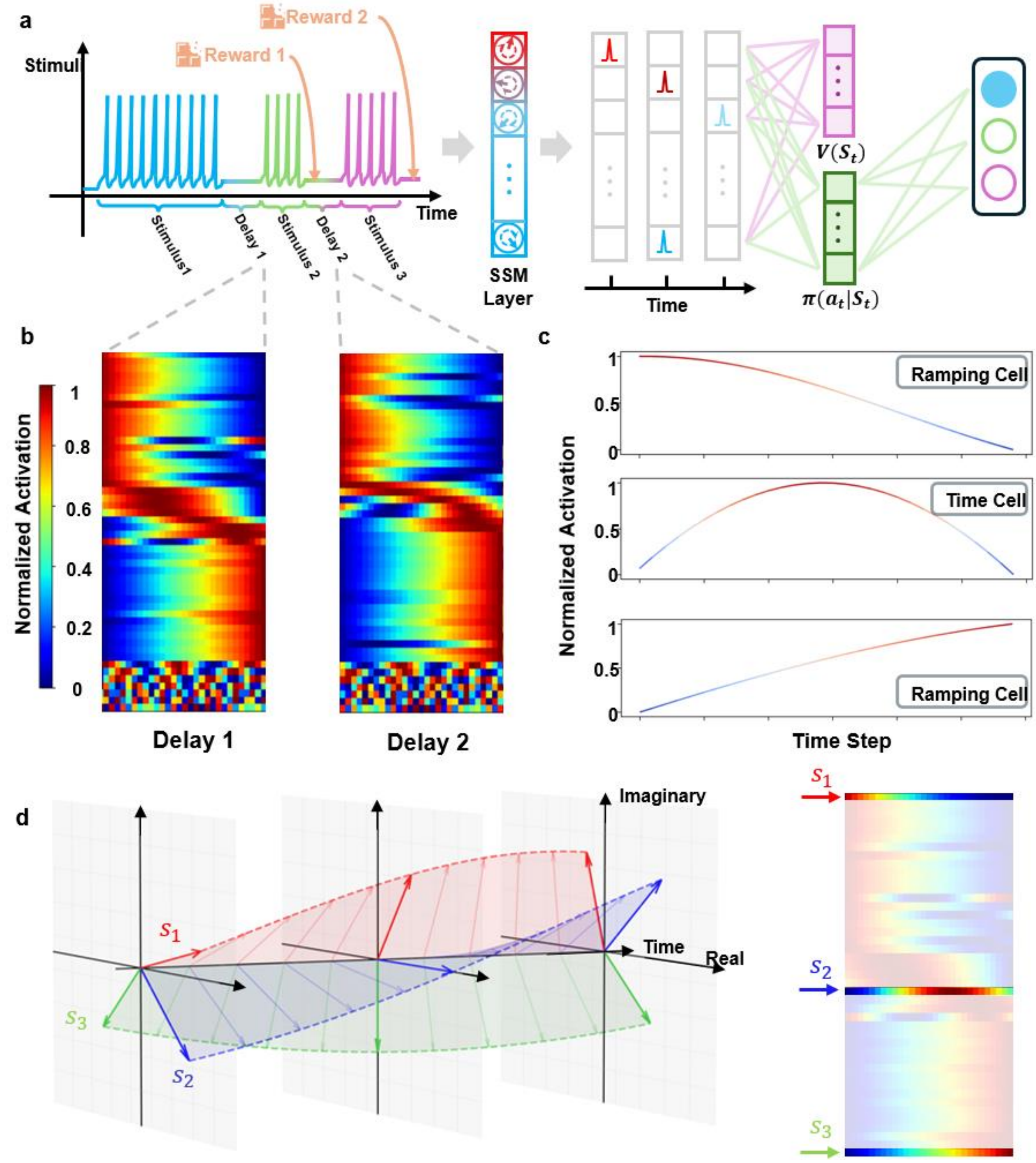

Figure 2 | Complex rotational dynamics naturally generate time cell behaviors in State Space Model (SSM). **a**, Learning Framework for the Temporal Judgment Task. The SSM layer processes three stimuli separated by two delay periods and is trained via reinforcement learning, with intermediate and final rewards, to identify the longest stimulus. On the right, the schematic illustrates the downstream readout architecture: continuous hidden states from the SSM layer are thresholded into discrete spiking signals. These spikes are then linearly projected to value network ($V(\mathbf{S}_t)$) and policy network ($\pi(\boldsymbol{a}_t|\mathbf{S}_t)$). **b**, Emergent Time Cell behaviors during Delay Intervals. Heatmaps of trial-averaged, normalized state activations during the two delay periods reveal diverse temporal coding mechanisms, with a few of them highlighted in **c**. **c**, Example Tuning Curves of Individual State Units. Detailed activation profiles of representative units classified as ramping cells (ramping up and ramping down) and time cells. **d**, A visual illustration of how rotations of 3 types of complex-valued states generate different temporal activation profiles through projection onto the real axis. $\mathrm{S}_1, \mathrm{S}_2, \mathrm{S}_3$

corresponds to ramping-down, time-cell-like, and ramping-up profiles respectively in our heatmap 2b.

While the trained SSM exhibits sophisticated temporal dynamics driven by complex rotations, a question remains regarding the origin of these representations: Are they constructed de novo through reinforcement learning, or do they arise from the network's intrinsic structure? To disentangle the contributions of structural inductive biases from learning-driven optimization, we compared the internal dynamics and performance of SSMs with HiPPO initialization versus random initialization.

As shown in Fig. 3a-b, both networks can learn the task with high proficiency, achieving comparable accuracies. However, their underlying dynamic solutions differ drastically. Visualizing the hidden state activations reveals that the HiPPO-initialized network exhibits sequentially organized time cell patterns even before training, a structure that is preserved and refined throughout the learning process without specifically training for it. This indicates that the HiPPO initialization equips the network with a set of preconfigured basis functions capable of representing temporal history. This computational observation aligns with recent physiological findings in the human brain organoid, where structured backbone firing sequences emerge intrinsically from neural architectures even without learning[27]. Further experiments on the variations of HiPPO initializations generalize the time cell behavior to the HiPPO family with complex eigenvalues and appropriate frequency distribution (see Supplementary Note 4, Supplementary Fig. 6 and Supplementary Table 2 for the full initialization ablation). On the contrary, even after extensive training, a randomly initialized network only produced unstructured, high-entropy dynamics with no discernible temporal order (Fig. 3a, right), suggesting that the SSM architecture alone with pre-configured basis functions is insufficient to support stable sequential representations.

To demonstrate the mathematical mechanism underlying this dichotomy, we quantified the plasticity requirements of different network components during training. We tracked the relative change of the trainable parameters—the state transition matrix ($\mathbf{\Lambda}$), the input matrix ($\mathbf{B}$), and the readout matrix ($\mathbf{C}$)—by computing the normalized distance between the initial and trained values ($|\theta_{final} - \theta_{init}| \ / \ |\theta_{init}|$). As shown in Fig. 3c, a clear functional dissociation emerges between the internal dynamics and the readout layer. For the internal state parameters, the HiPPO-initialized network exhibits minimal change. The relative changes for $\mathbf{\Lambda}$ (mean $\approx$ $0.093 \pm 0.003$) is significantly lower (12.4×) than that in the randomly initialized network ($\mathbf{\Lambda}$ mean $\approx$ $1.154 \pm 0.038$). It is worth noting that the temporal nature of the tasks may lead the network to prefer temporally expressive initialization. Nevertheless, structural stability confirms that HiPPO initialization provides a near-optimal encoder structure in which the input information is compressed into the hidden state using preconfigured basis functions, requiring only minor fine-tuning rather than extensive restructuring.

In contrast, we also observed that $\mathbf{B}$ and $\mathbf{C}$ exhibit relatively larger changes for both HiPPO ($\mathbf{B} \approx 0.997 \pm 0.001, \mathbf{C} \approx 2.232 \pm 0.278$) and randomly initialized SSM ($\mathbf{B} \approx 1.134 \pm$ $0.058, C \approx 1.824 \pm 0.089$) at roughly equal rates (1.1~1.2×). Furthermore, since both networks initialize $\mathbf{C}$ randomly, they must learn to extract task-relevant information from the hidden states. This suggests that in the HiPPO-initialized SSM, training is largely through re-configuring connections of the basis functions to neurons in the downstream readout layer, while in a random-initialized network, training involves significant updating of all parameters throughout the network. The latter case can still produce desired final outputs after aggressive training, but fails to explain the role of time cells observed in biology.

This structural hypothesis is further validated by the parameter-freezing experiment. Specifically, when we froze the internal parameters ($\mathbf{\Lambda}$ and $\mathbf{B}$) to their initial values and only trained the readout layer, the HiPPO-initialized network maintained an accuracy of $50.4 \pm 3.5\%$. As noted in the initial HiPPO work, training B produces minor but consistent improvement[35], which is consistent with our results. On the other hand, the random-initialized network's accuracy collapsed to approximately $34.1 \pm 2.3\%$ under the same conditions. With internal dynamics fixed, the random-initialized network's state representation carries insufficient temporal information for the readout to recover, while the HiPPO basis routes that information to the readout in a directly decodable form. Task performance and the emergence of time cells thus depend substantially on the quality of the basis rather than on plasticity alone.

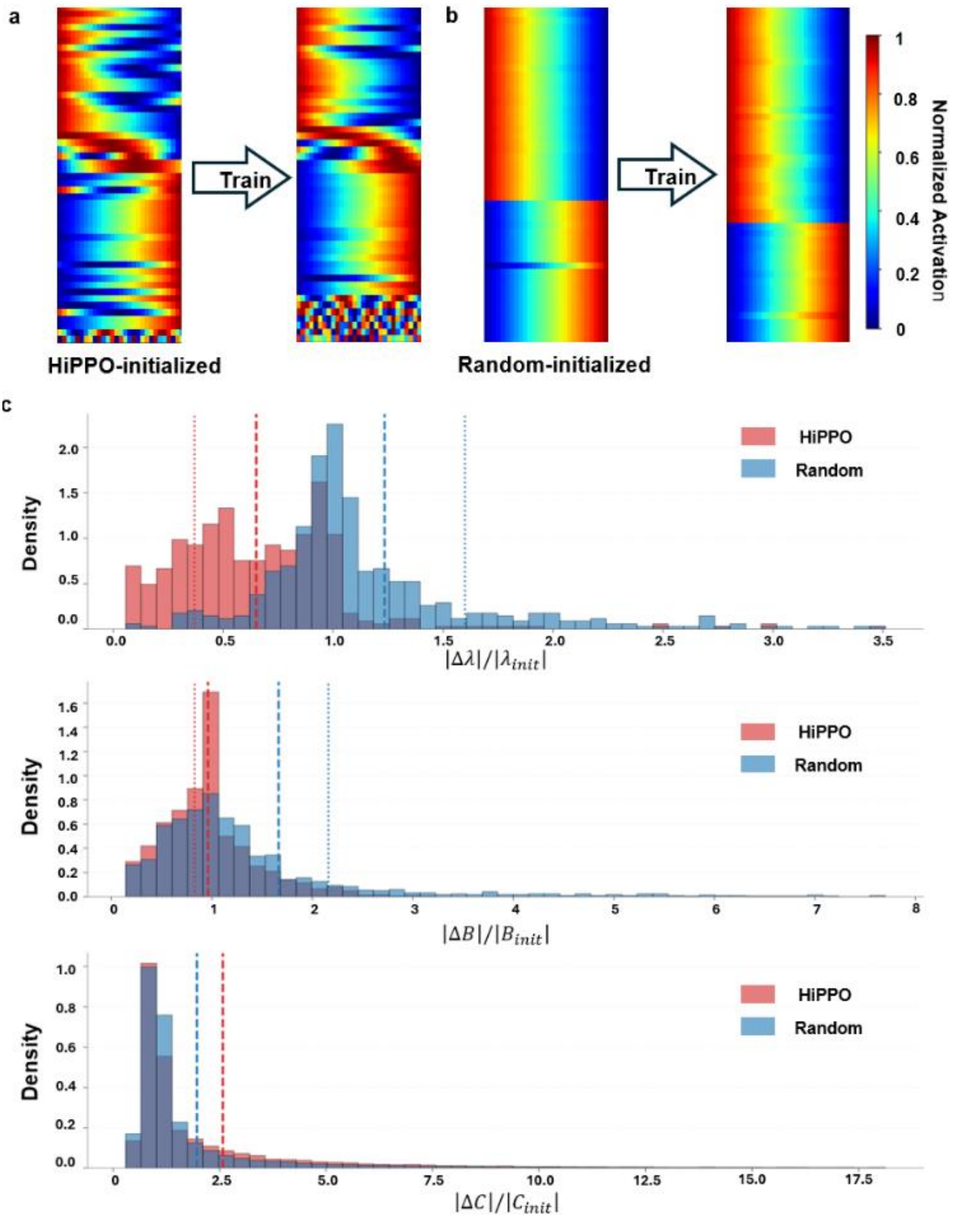

Figure 3 | Optimal initialization provides a robust structural basis, shifting the learning burden from internal dynamics to the readout layer. High frequency noises are removed for clearer presentation **a** and **b,** Intrinsic dynamics of untrained networks. Heatmaps of hidden state activations show that HiPPO-initialization generates time cell behaviors prior to training, whereas random initialization produces unstructured noise. **c**, Quantification of parameter plasticity. Histograms showing the distribution of relative parameter changes ($|\Delta\ \theta|\ /\ |\theta_{init}|$ ) after training with vertical dashed lines indicating the mean values. In the HiPPO-initialized network (red), $\mathbf{\Lambda}$ exhibits minimal changes

compared to the Random network (blue). For the input adaptation matrix **B** and the output matrix **C**, the HiPPO-initialized network exhibits a larger relative change, demonstrating that the preconfigured state space allows the model to solve the task primarily by optimizing the input projection and downstream readout while preserving the oscillatory basis.

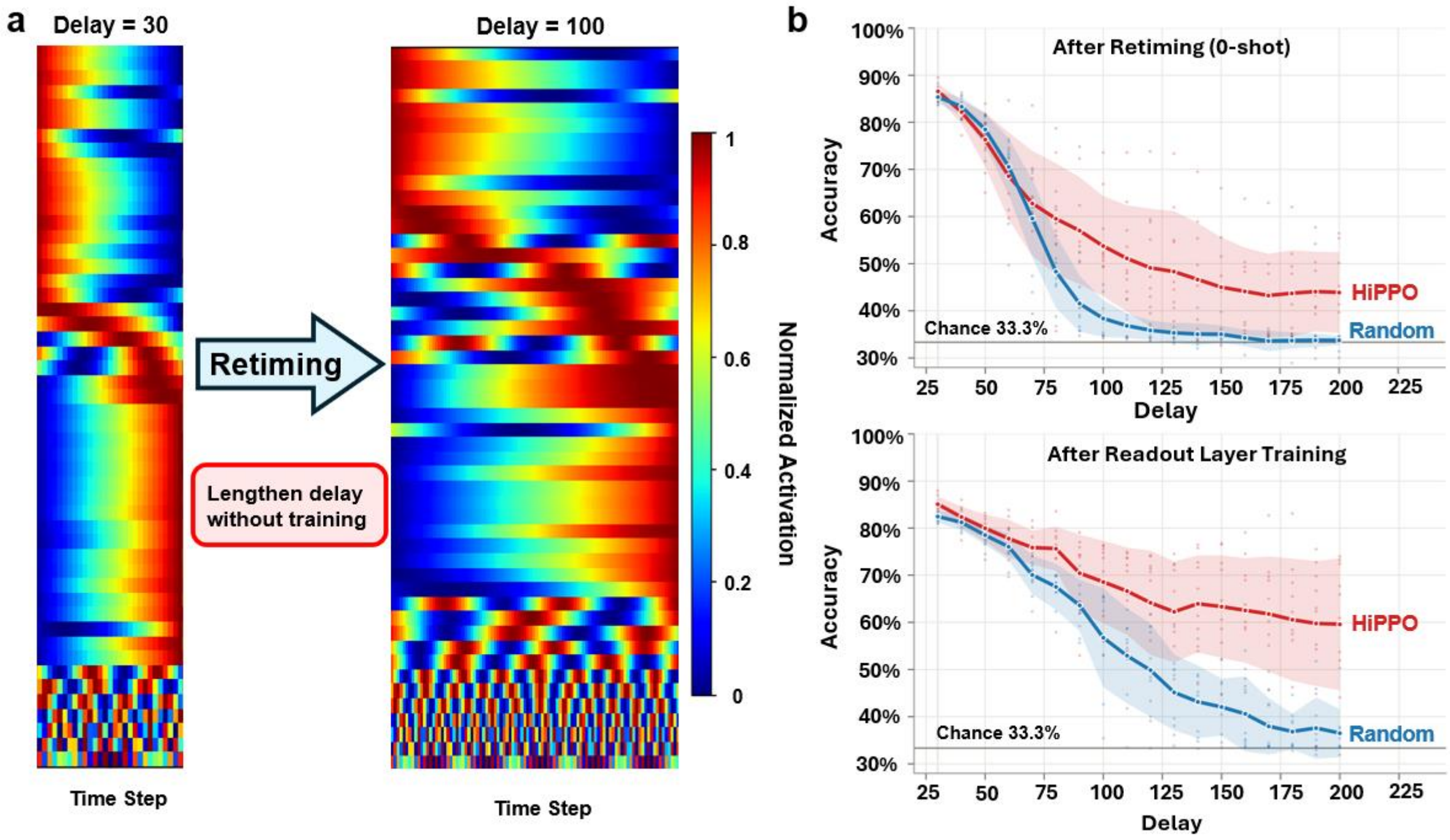

Figure 4 | Intrinsic Scalability of Temporal Representations. **a**, Heatmaps show that a network trained on a 30-step delay naturally produces stretched time-cell sequences when the delay is extended to 100 steps via retiming, even without retraining internal parameters. **b**, Divergent recovery of performance between initializations. While directly extending the delay causes a drop in accuracy, retraining only the linear readout layer (while keeping internal dynamics **Λ** and **B** frozen) restores the performance to some extent. In contrast, the randomly initialized network fails to adapt after a longer delay period, despite starting with comparable performance. This demonstrates that the learned oscillatory basis is robust and scalable, requiring only superficial re-calibration to adapt to new timescales. Error bands are plotted at $\pm 1$ standard deviation.

## Retiming and Parameter-Freezing Experiment

To test whether the proposed framework captures biological adaptability, we performed a retiming experiment. Here, an agent originally trained on a short (30-step) delay was assessed on a significantly extended (up to 200-step, but we use 100-step as an example) interval without any retraining, to observe intrinsic scaling properties following Mello et al.[18]. The new internal state dynamics are shown in Fig. 4a, where we simply extended the delay period in the input stream while keeping all network parameters frozen at their values learned from the 30-step task. As evidenced by the heatmaps, the learned time-cell sequences did not reset; instead, the extended delay in the input allowed the pre-existing oscillatory modes to continue their respective rotations, with their projections on the real axis effectively tiling the entire 100-step interval. Notably, the sequential order in Fig. 4a (right) was obtained by re-sorting the units in the extended delay. We found that the unit indices between the two conditions are not one-to-one corresponding: a unit coding for a specific relative phase in the short delay does not necessarily encode the same relative phase in the long delay. This test shows that the rotational dynamics provide a scalable temporal substrate, mirroring the intrinsic population-level remapping observed in the hippocampus[20,21]. Note that the advantage of HiPPO initialization gets more significant as the delay duration extends, as illustrated in Fig. 4b. We hypothesized that the network only required a superficial recalibration. Indeed, by retraining only the readout-related layer while keeping the internal state dynamics (**Λ** and **B**) completely frozen, the network's accuracy rapidly restores. The corresponding LSTM recovery comparison is shown in Supplementary Fig. 5. The same trend is also observed in the baseline models trained with 60-step delay (Supplementary Note 5 and Supplementary Fig. 7).

Similarly, we performed the same parameter-freezing experiment on the randomly initialized network. Although the randomly initialized network solved the original 30-step task with equivalent high accuracy, its performance reduced to chance-level for the extended delay (175+ steps) and, unlike the HiPPO-initialized network, failed to recover after readout retraining (Fig. 4b). These results further support the hypothesis that the optimally initialized SSM possesses sufficient mechanisms to process time information that can be reused across timescales, qualitatively consistent with what has been observed in biological systems.

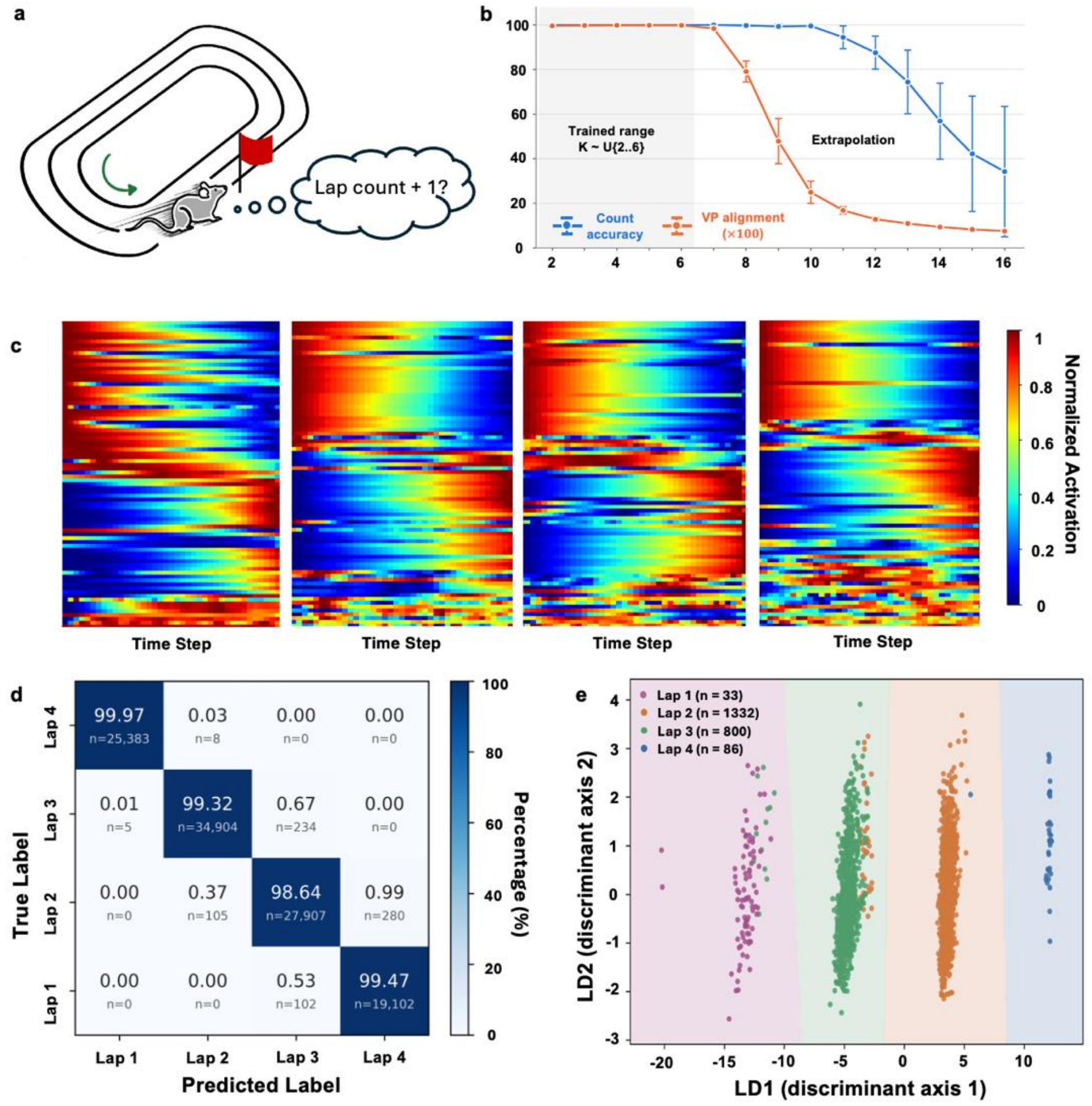

Figure 5 | Generalization of Temporal Coding to a Cyclical Lap-Counting Task **a**, Conceptual Diagram of the Lap-Counting Task. An agent processes a repeating sequence with a landmark and must learn to increment a counter after each lap. **b**, Accuracy vs Victor-Purpura (VP) score Performance. The network's predicted lap count and VP-score (higher score indicates accurately increment the lap count near lap ending signal) over time shows near perfect performance in-distribution. Error bars are plotted at $\pm 1$ standard deviation. **c**, Re-emergence of Time Cell Sequences Across Laps for 4 laps. The sequential activation of time cells resets and reliably replays during each consecutive lap. The instances fluctuate because of random variations. **d**, The Lap-Wise Separability Confusion Matrix. **e**, States projected onto the first two linear-discriminant (LD) axes (LD1 and LD2) through Linear Discriminant Analysis (LDA). Shaded regions are the decision regions of a linear support vector classifier (SVC) on four laps trained on the same data points.

## Lap-Counting Experiment

Finally, we show that the same framework can be used to explain more abstract cognitive behaviors by simply connecting the internal SSM states to other downstream networks. To challenge the proposed model beyond simple interval timing and test the generality of the timing mechanism, we designed a more abstract lap-counting task. As conceptualized in Fig. 5a, the agent processes an input stream representing two alternating types of events: (1) A running sequence, which represents the agent's progress along the track. (2) A distinct finish line landmark sequence. This experiment is inspired by the studies on mouse brains in Sun et al.[39] , where hippocampal neurons were found to represent discrete, repeated lap events as fundamental units of an experience.

As discussed in Sun et al.[39], this is a much harder challenge for the agent as the agent must compress the temporal information of running (with identical sensory inputs in different laps) and specifically recognize the landmark sequence as the definitive cue to increment its internal lap count, demonstrating a move from simple timekeeping to abstract event recognition (detailed training description in Supplementary Methods 1.2 and 1.3).

The network achieved 100.0% accuracy over 4 laps for the lap-counting task. However, accuracy alone is not informative, since the network could also have correctly counted the laps at incorrect timing, with a random chance of correctly firing at a landmark of 0.30% chance (obtained by running a random untrained network). We therefore employed an edit-distance metric to measure the dissimilarity between the predicted lap-count timings and the ground-truth timings to evaluate our model. This score is calculated by computing the cost of editing the predicted sequence to match the ground truth sequence. The proposed trained network achieved a normalized edit distance score (see Supplementary Methods 2, Supplementary Equations (1) and (2)) of 0.999 (maximum of 1.0 for the exact same timings), indicating high alignment. Fig. 5b shows the network's average performance within the training distribution and extrapolated beyond 6 laps, each lap with a random duration ranging from 10 to 60 steps and 0~2 random pauses of 0 to 25 steps. The VP score indicates the model’s performance extending to more laps before reaching its limit.

Similarly, the analysis of the hidden-state dynamics indicates that the model naturally achieved this through reward, without any explicit instructions for temporal coding. The heatmap in Fig. 5c visualizes the normalized activation of the hidden state units. Similar to Figs. 2-4, the units were independently sorted by their peak activation time within each of the four laps. The result shows a sequential cascade, the signature of time cell behavior, reliably re-emerged in every single lap (Fig. 5c). The intuitive question here is then: if the temporal firing patterns reset for each lap, how does the system maintain the persistent count information required to distinguish Lap 1 from Lap 2?

To investigate whether this abstract cognitive variable is explicitly represented in the neuronal dynamics, we performed a decoding analysis using a linear Support Vector Classifier (SVC). Because lap durations, pause durations, and the number of laps per episode are all randomized, different episodes are on different laps at the same absolute time, we can only decode at a specific timestep, so every sample entering a given classifier shares an identical elapsed time from episode onset and temporal position cannot contribute to the decode. We extracted the hidden-state activity from the stable step of each episode and tested its separability. To test the robustness of the representation, we trained the classifier with a 50/50 split (train / test).

As shown in the confusion matrix in Fig. 5d, the linear classifier achieved near-perfect performance on the unseen test set. This result shows that, despite receiving highly similar sensory streams, the high-dimensional internal state for Lap 1 is distinct from that for Lap 2. The fact that a linear classifier could solve this task with minimal training data implies that the abstract concept of count is not entangled in some complex manifold but appears to be linearly separable.

This geometric organization is visualized in Fig. 5e, where the decision boundaries are projected onto the first two linear discriminant axes and fitted to a support vector classifier. The state space is partitioned into distinct regions corresponding to each lap count. This finding is significant because it suggests that the HiPPO-initialized SSM naturally unrolls repetitive sensory inputs into a structured trajectory. The recurrent dynamics automatically segregate repeated events into distinct locations in the state space, allowing downstream readout layers to extract abstract variables such as event count via simple linear projections, without requiring additional non-linear processing.

## DISCUSSION

Our results demonstrate that sequential time-cell-like behaviors can emerge from a preconfigured structure for compressing a continuous history. By analyzing SSMs initialized with HiPPO dynamics, we propose a framework that offers a route to reconciling the millisecond timescale of cellular dynamics with the behavioral timescale of working memory. The rotational dynamics inherent in the HiPPO framework are necessary for orthogonal polynomial projection and they qualitatively correspond to biological oscillatory activity. This correspondence raises the possibility that the brain's intrinsic dynamics instantiate such basis functions. We note that the proposed framework is a mathematical model that offers mathematical sufficiency and computational plausibility to generate response profiles resembling time cells, ramping activity, and oscillatory responses, but not a proof that biological evolution produced an equivalent basis. Additional studies that perform direct neural-data comparisons or through biologically constrained circuit models are needed to make stronger connections to neural circuits.

The HiPPO-initialized SSM utilizes a diagonal state transition matrix with complex-valued coefficients. While this design is primarily a computational choice, it is worth noting that it is not inconsistent with known constraints on biological oscillators. Studies on the reduction of the 4-dimensional Hodgkin-Huxley model demonstrate that while the system can be simplified, it requires at least 2 dimensions to retain the capacity for oscillation (e.g., sub-threshold oscillations)[40]. A one-dimensional biophysical model is over-damped and incapable of resonance. Our use of complex-valued states, which mathematically represent conjugate pairs of real variables, satisfies this minimal dimensionality requirement. Two-dimensionality is a necessary condition for oscillation, which is met by many distinct circuit implementations. The rotation in our state space is an abstraction of oscillatory dynamics at the computational level. Hence, it is consistent with the possibility that time cells and other temporal codes reflect a preconfigured structure that learning adapts.

Finally, the oscillatory dynamics that our framework relies on have qualitative counterparts across multiple scales of neural organization. At the population level, the reorganization of our model under retiming is qualitatively comparable to remapping reported in hippocampal circuits. During rhythmic timing tasks, neural populations in the supplementary motor area generate stable, periodic state-space trajectories[41–43], and related work suggests distributed neural dynamics within the striatum and pre-SMA in voluntary motor control[44–46]. Together, these observations suggest that rotational dynamics are a recurring motif in circuits governed by timing, making the present framework a plausible computational description of such systems.

Several limitations of the current study still need to be addressed to fully contextualize these findings. First, the complexity of the tasks employed may not fully exploit the capacity of the proposed model. Although by strategically minimizing structural complexity, we can isolate the specific contribution of rotational dynamics to temporal processing, ensuring that the observed emergent behaviors are driven by the proposed computational principles. However, the fact that randomly initialized networks achieved relatively high accuracy suggests that the chosen tasks might be solvable with naïve designs, rather than requiring the robust, long-term history compression provided by HiPPO initialization. Additionally, we also observed high variance in performance among different seeds and hence the performance that we reported could be lower than the best models tuned under this framework. Second, while HiPPO represents a mathematically optimal solution for online function approximation, it is likely only one of a class of suitable initializations, and there is no biological guarantee that the brain utilizes this specific orthogonal basis over other potential pre-configured dynamics. Finally, our model abstracts away the intricate microarchitecture of biological circuits. While the hippocampus exhibits complex connectivity patterns and diverse cell types, our model relies on a simplified state transition structure without such heterogeneity. At the physical level, these rotational dynamics may be instantiated as intrinsic node-level properties within individual neurons, providing a potential hierarchical framework.

One potential future direction that these limitations point toward is a fertile intersection between control theory and evolutionary biology, such as the domain of meta-learning[47,48]. If evolution is viewed as a grand optimization process, the intrinsic dynamics observed in biological circuits could be interpreted as a meta-learned initialization. Future work should investigate whether evolving a population of SSMs over generations, selecting purely for memory capacity, causes weight matrices to spontaneously converge toward HiPPO-like structures, thereby testing whether mathematical optimality acts as an attractor under selection pressure for memory. Additionally, while we have established that rotational dynamics generate temporal representations in our model, we did not explicitly analyze the distribution of these frequencies. Future work could systematically quantify the spectrum of these learned rotational frequencies and compare them to specific biological oscillation bands (e.g., theta, gamma) to test whether the proposed mathematical framework aligns with additional biophysical observations. Finally, the current iteration of the proposed model relies on artificially encoded input signals, which simplifies the noisy nature of biological information transmission. Since we have established a proof-of-concept linking continuous state-space dynamics to time-cell-like activity, it will be critical to validate this framework using real spiking data, challenging the model to maintain robust historical representations amidst the sparsity and stochasticity inherent to actual neural substrates.

In conclusion, our study suggests that the simple principle of rotational dynamics may provide a sufficient computational basis for time perception. The SSM-based model reproduces several classes of biological temporal response profiles within a single set of computational principles and provides a mathematical framework for generating testable hypotheses about the algorithmic basis of temporal cognition. This approach contributes to the ongoing effort to develop a multi-scale theory of how the brain learns to operate in time and offers a mathematical perspective on how neural circuits might function.

# METHODS

## Classification of Functional Cell Types

To characterize the temporal coding properties of hidden state units in the trained network, we analyzed their trial-averaged activity during the delay periods. Following successful training, we recorded the hidden state activations across 1,000 episodes with the network weights frozen. For each unit, we computed a trial-averaged temporal tuning curve. Based on the profile of this curve, units were classified into time cells and ramping cells.

Our criteria for identifying time cells and ramping cells were adapted from Shikano et al.[49].

- Ramping Cells**:** A unit was classified as a ramping cell if its temporal tuning curve showed a monotonic increase or decrease. To quantify this, we fitted a linear regression to the tuning curve. A unit was considered a candidate ramping cell if the fit yielded a significant p-value ($p \leq 0.05$) and a high Pearson correlation coefficient ($\geq 0.9$).
- Time Cells**:** A unit was classified as a time cell if it fired selectively at a specific moment in the delay period. We identified these by calculating the temporal information of the tuning curve based on the Skaggs et al.[50] formula. If a unit was already a ramping candidate, its temporal information was calculated from the tuning curve after subtracting the linear regression fit to isolate peak-like activity from the ramp. A unit was considered a candidate time cell if its temporal information was significant (i.e., higher than the 99th percentile of a distribution generated from 100 shuffled tuning curves).

To ensure the temporal profiles were stable and not due to chance, all candidate ramping and time cells were subjected to a trial-reliability analysis. We computed a reliability score by correlating the tuning curves obtained from even- and odd-numbered trials. Only candidates with a significant reliability score (higher than the 99th percentile of scores from 100 shuffled comparisons) were confirmed as actual ramping or time cells. Note that these classifications were not mutually exclusive; a single unit could meet the criteria for multiple categories.

## Sorting Into Heatmaps

Given the broad frequency spectrum inherent to HiPPO initialization, we applied a two-stage sorting algorithm to isolate task-relevant temporal features for visualization. First, raw state trajectories were range-normalized and categorized by their oscillatory frequency, defined as the count of local maxima exceeding a specific amplitude prominence threshold. Second, within each frequency group, units were ordered by peak latency. Notably, units displaying one or two peaks were aggregated into a single cluster to account for boundary effects, where low-frequency time cells may exhibit high values at both temporal edges. Implementation details are provided in the code repository.

## Actor-Critic RL Framework

Our framework, inspired by the design of Lin et al.[17], employs an actor-critic architecture with an SSM memory module. The SSM processes the raw input signal stream to generate a state representation, which is fed to both the actor and critic networks.

The actor-critic architecture is a widely used framework in reinforcement learning[33] that integrates both policy-based and value-based methods. It consists of two components: an actor that learns a policy to select actions based on the current state, and a critic that estimates a value function to evaluate the quality of actions or states. The agent was trained to minimize the loss of the policy and value networks, $L = L_\pi + L_V$, where

$$L_\pi = \sum_{t=0}^{T-1} -\log \pi * (R_t - \hat{V}) \tag{4}$$

$$L_V = \sum_{t=0}^{T-1} l_1(\hat{V}, R_t) \tag{5}$$

where, $T$ is the total timesteps of the trial in the episode, and $R_t = \sum_{i=0}^{t} \gamma^i r_{t-\mathrm{i}}$ is the reward computed at time $t$ based on past reward, $\gamma$ is the discount factor, and $l_1(\cdot)$ is the smoothed L1 loss.

The critic network, implemented as a linear layer, estimates the scalar state value, $\hat{V}(\mathbf{S_t}, \boldsymbol{\theta})$. The actor network, also a linear layer, outputs a policy, $\pi(a_t|\mathbf{S_t}, \boldsymbol{\theta})$, which is a probability distribution over the action space. A discrete action is then sampled from this distribution using a Softmax function with a temperature of 1. At the beginning of each trial, the SSM's hidden states are initialized to zero.

Throughout our experiments, we employed the actor with 1 neuron and a critic with 1 neuron fully connected to the SSM output. The list of hyperparameters has been summarized in Supplementary Methods 3 and Supplementary Table 3.

During training, the actor updates its policy in the direction suggested by the critic, using the temporal-difference (TD) error based on the discounted cumulative reward, which captures both immediate and future rewards with a decay factor, as a learning signal. On the other hand, the critic minimizes the TD error to improve the accuracy of its value estimates. This cooperative interaction enables stable and efficient policy learning. During inference, only the actor is used to generating actions, as the critic is not required once the policy has been learned. In particular, the final action is sampled based on the probability of selecting an action predicted by the policy network.

## DATA AVAILABILITY

Source data are provided with this paper. The data used in this study are generated in real time during training and inference under the GitHub repository (see CODE AVAILABILITY). The experimental result data generated in this study are provided in the Source Data file.

## CODE AVAILABILITY

The codebase is publicly available at https://github.com/lusensama/SSM_timecell.git under MIT license[51]. It contains the environments generating the training and evaluation data, training scripts, analysis scripts used to produce figures and tables in the manuscript and Supplementary Information.

## ACKNOWLEDGMENT

W. D. L acknowledges support from the James R. Mellor Professorship.

## FUNDING STATEMENT

This work was supported in part by the National Science Foundation under Grant CCF-2413293 and by the Michigan Semiconductor Talent and Technologies for Automotive Research program.

## AUTHOR CONTRIBUTION STATEMENT

W.D.L conceived the project. S.L. and X.Z. designed the experiments. S.L. and X.Z. performed model training. W.D.L, S.L, X.Z., M.H., E.Y.L., S.K. contributed to the discussion and contributed to the manuscript writing at all stages.

## COMPETING INTERESTS

The authors declare no competing interests.

# Supplementary Information

## Supplementary Methods 1 Task environments

### 1.1 Stimulus discrimination task

In this task, the agent is presented with three random-length stimuli (sampled from [10,15, 20,25, 30,35, 40,45, 50] without replacement) and must decide the longest stimulus. A correct final prediction at the end of the trial yields a reward of +10, while an incorrect one results in a penalty of -10. Additionally, an intermediate reward or penalty of ±5 is given based on the prediction for the first two stimuli only. The inputs are one-hot coded, and the delays are all zeros.

### 1.2 Lap counting task

The agent runs laps of a notional circular track and must emit one pulse per completed lap. The observation is two-dimensional and one-hot coded:

- [1, 0] running
- [0, 1] lap-end landmark (a single timestep)
- [0, 0] paused, or post-episode padding -- the EMPTY signal, no input at all

The action space is binary: 0 = do nothing, 1 = emit one pulse into the lap counter.

Reward. The reward is a function of the observation. It receives whether a landmark was just observed, what the agent did, and the state of the response window (itself derived from landmark observations). Specifically:

- first emission within hit_window steps AFTER an observed landmark +1.0
- a landmark's response window closes with no emission -1.0
- emission outside any open window (false alarm) -0.1
- second or later emission inside the same window -0.1
- terminal: emitted lap count same as true lap count +/-2.0

Only the FIRST in-window emission increments the counter, so the reported count is exactly one pulse per acknowledged lap and is decoupled from press noise; extra emissions are penalized but do not increment.

Because of that increment rule, count accuracy alone is not a sufficient measure: a policy that emits at every timestep also reaches 100% count accuracy, at the cost of ~650 false alarms per episode. Count accuracy is therefore always reported together with response alignment (Victor-Purpura distance, tolerance 3 steps, q = 2/tolerance).

Task chance was measured by running naive policies through the environment rather than computed from a binomial expression: never emitting and emitting a fixed number of pulses at episode start both give 0% count accuracy, i.i.d. emission at the landmark rate 0.30% (18.5 false alarms/episode), and i.i.d. emission at p = 0.05 gives 1.10%.

### 1.3 Randomized runs and training

To prevent the agent from timing rather than detecting laps, every temporal regularity of the task is randomized per episode:

- lap count K ~ Uniform{2, ..., 6}
- lap duration Uniform{10, ..., 60} running steps, drawn INDEPENDENTLY per lap
- within-lap pauses 0-2 per lap, each Uniform{0, ..., 25} steps, inserted at INTERIOR running-step boundaries of that lap
- episode length every episode padded to exactly 666 timesteps (maximum laplengths possible)

A pause is the empty observation [0,0], i.e. the absence of input rather than an uninformative constant. A pause after a landmark would only lengthen the interval between landmarks, whereas a mid-lap pause breaks the monotone correspondence between position along the lap and elapsed time within the lap.

Training details: A2C, 80 recurrent units, learning rate 3e-3, weight decay 1e-6, gamma 0.9, entropy coefficient annealed 0.1 -> 0.01, 20,000 episodes per seed, 3 seeds, on CPU. Further implementation details can be found in the published codebase.

## Supplementary Methods 2 Normalized VP score and edit distance

Edit distance is a common way to measure the correctness of timing. A normalized Victor-Purpura (VP) score for the predicted and ground-truth timings is computed in two steps. First, it finds the cost to edit the prediction into the ground truth

$$\mathrm{D}(i,j) = \min \begin{cases} \mathrm{D}(i-1,j) + 1 \\ \mathrm{D}(i,j-1) + 1 \\ \mathrm{D}(i-1,j-1) + q \cdot |p_i - g_j| \end{cases} \quad (1)$$

In Supplementary Equation (1), the three terms are the costs of deleting a predicted event, inserting a ground-truth event and shifting a predicted event to a ground-truth event, respectively.

And the final edit distance is the last cell $d_{\mathrm{VP}}(\mathbf{P}, \mathbf{G}; q) = D(m, n)$.

Second, it normalizes the distance against the maximum possible distance (deleting all predictions and inserting all ground truth), the normalized VP score is given by:

$$\mathrm{Score}_{\mathrm{VP}}(\mathbf{P}, \mathbf{G}; q) = 1 - \frac{\mathrm{d}_{\mathrm{VP}}(\mathbf{P}, \mathbf{G}; q)}{|\mathbf{P}| + |\mathbf{G}|} \quad (2)$$

where, $d_{\mathrm{VP}}(\mathbf{P}, \mathbf{G}; q)$ is the edit distance, $|\mathbf{P}|$ is the number of events in the prediction list, $|\mathbf{G}|$ is the number of events in the ground truth list, and $q$ is the cost for shifting. We define this normalized VP score as the normalized edit distance in the manuscript.

## Supplementary Note 1 Lesion study

We conducted lesion studies that progressively removed hidden-state units and assessed the impact on the stimulus discrimination task. We evaluated three additional independently trained models alongside the model used in the main text. Each model was evaluated over 25,000 independent random trials at each lesion level. Supplementary Fig. 1 compares ascending and descending unit-index lesion orders; Supplementary Fig. 2 shows the single-unit lesion results for the model used in the main text.

The lesion study demonstrates an asymmetry in performance degradation that aligns with the spectral organization observed in Supplementary Fig. 3. In the forward lesion order (blue curve), units are removed starting from the lowest indices (1 to 50). As shown in Supplementary Fig. 3, these low-index units correspond to high-frequency oscillatory modes. The network maintains high accuracy even after removing the first ~20 of these units, suggesting that high-frequency modes are less critical for this temporal discrimination task. In stark contrast, the reverse lesion order (orange curve) targets the high-index units first (50 to 1). Since these units correspond to the low-frequency modes such as time cell and ramping dynamics identified in Supplementary Fig. 3, their removal causes an immediate and precipitous drop in accuracy. This suggests that the model's ability to bridge delay periods relies on the slow-rotating basis functions located at the higher indices of the HiPPO spectrum[1].

Additionally, the single unit lesion study also does not show any significant degradation for any particular units and the units causing the most accuracy drop in this figure ($35^{th}$ and $40^{th}$) are not involved in the cascading of the time cell pattern in SSM (see main text, Fig. 2b).

## Supplementary Note 2 Individual hidden state tuning curves

To supplement the population-level analysis in Fig. 2c of the main text, Supplementary Fig. 3 shows the complete set of trial-averaged, normalized and unsorted activation profiles for all 50 hidden state units from a representative trained model. The two panels correspond to the first and second delay periods.

Supplementary Fig. 3 displays the trial-averaged, normalized activation profiles for hidden state units during the delay period. A candidate cell is confirmed only if the tuning curves computed from even- and odd-numbered trials correlate above the $99^{th}$ percentile of 100 shuffled comparisons, and time-cell candidacy separately requires temporal information above the 99th percentile of 100 shuffled tuning curves. The organization of these temporal dynamics aligns with the theoretical prediction of the HiPPO initialization[1], where the rotational frequency of a state unit (determined by $\lambda_i$) is inversely proportional to its index ($n$). Consequently, units with lower indices (visible in the earlier panels) retain significantly high-frequency oscillatory components. As the unit index increases, the rotational frequency systematically decreases, naturally giving rise to the slower, smoother, and more persistent dynamics characteristic of time cells and ramping cells. This index-dependent spectral distribution confirms that the diverse temporal receptive fields observed in the population are not random, but are derived from the pre-configured, multi-scale rotational structure of the SSM.

## Supplementary Note 3 Comparison with an LSTM network

To benchmark the performance of the State-Space Model (SSM) and investigate the uniqueness of its emergent temporal coding, we conducted a comparative experiment using a Long Short-Term Memory (LSTM) network, a standard architecture for sequence processing. We replaced the S5 SSM layer in the model with an LSTM layer, while the rest of the actor-critic reinforcement learning (RL) framework, including the downstream policy and value networks and the actor-critic training algorithm, remained identical to that used in the main study. The LSTM-based agent was trained on the same three-stimuli temporal discrimination task.

It is worth mentioning that the LSTM network used in the previous study[2] took more effort to train for the three-stimuli task, despite achieving perfect accuracy for the two-stimuli task. Supplementary Fig. 4 demonstrated the timecell cascade that is taking shape, consistent with original work.

However, Supplementary Table 1's comparison between the LSTM and SSM demonstrate that the SSM outperforms LSTM under the same compute budget.

Supplementary Fig. 5 also shows that freezing the internal dynamics basically freezes the model performance and retraining only the readout layer can't recover the performance.

These results suggest that while LSTMs are powerful sequence models, their internal mechanisms do not inherently or spontaneously give rise to the oscillator-based, time-tiling representations but are produced through learning. This highlights the architectural advantage of the SSM framework for solving tasks that require precise temporal representation and discrimination.

## Supplementary Note 4 Initialization ablation studies

We have designed additional tests in an attempt to isolate a property of HiPPO-initialization by holding everything else at its original HiPPO initialized structure:

- spectrum_matched: the eigenvalue multiset is exactly HiPPO's, but the imaginary parts are randomly permuted
- freq_matched: imaginary parts exactly HiPPO's, decay rates randomized
- perturbed_hippo: HiPPO plus noise (eps = 0.1), potentially inflates the coverage score
- alt_basis: a different structured basis over the same frequency range (but evenly separated)
- s4d_lin and s4d_inv: the S4D-Lin and S4D-Inv initializations in Table 3 of ref. 1, defined by Supplementary Equations (3) and (4), respectively.
    - s4d-lin: $\mathbf{A}_n = -\frac{1}{2} + i \cdot \pi \cdot n$ (3)
    - s4d-inv: $\mathbf{A}_n = -\frac{1}{2} + i \cdot \left(\frac{N}{\pi}\right) \cdot (\frac{N}{2n+1} - 1)$ (4)
- real_diagonal: multi-timescale but non-rotational: purely real eigenvalues, Re = -logspace(-2, 0), i.e. leaky integration
- rand_complex: the naive baseline of the submitted manuscript

Quantifying Timecell. In addition to accuracy, we quantify time-cell behavior using a simple metric to make results easier to read without needing to see the heatmap. Since time-cell behavior is the cascade of activation in the hidden state, we quantify it as the fraction of delay bins that are the peak of at least one active unit, i.e. whether the units tile the interval. It is measured on the untrained model and again after training, averaged over 10 seeds and both delay periods.

From Supplementary Table 2 and Supplementary Fig. 6, S4D-Inv is an independent confirmation, as it is defined solely to approximate the asymptotics of HiPPO's spectrum but not its construction, and it matches HiPPO's values for both properties (coverage 0.437 vs 0.452; accuracy 88.7 vs 88.1). These results show that cascade generation and learnability have different sufficient conditions. As shown by real_diagonal and rand_complex perform the task at ~87% accuracy, but everything peaks in ~2.5 of 30 bins. This indicates that the randomly initialized models did not solve the task via time-cell decoding. On the other hand, while alt_basis has the strongest diverse coverage in timesteps, its uniform frequency interval with naïve coverage does not optimally compress the history states, hence leading to poor performance. We note that our standard HiPPO has neither the best performance observed nor the best coverage, but it is also the most stable one (least variance). This generalizes overall effectiveness of the HiPPO-based initializations. A structured multi-timescale complex rotational basis is what's responsible for generating time-cell cascades and supports its decoding simultaneously.

## Supplementary Note 5 Retiming from a longer training delay

We have extended the retiming experiment to models trained on 60-delay and similarly extended to 200 timesteps. As shown in Supplementary Fig. 7, the HiPPO-initialized model consistently outperforms the randomly initialized model, especially when the model extends beyond 100 steps from the initial delay durations, where the randomly initialized model collapses to chance-level performance.

While both HiPPO and random suffer from accuracy degradation when retimed to long durations, HiPPO maintains decent performance even before retraining the output layer. This result shows that our conclusions in the main text are not limited to a single data point and can be generalized to an upscaled task.

## Supplementary Figures

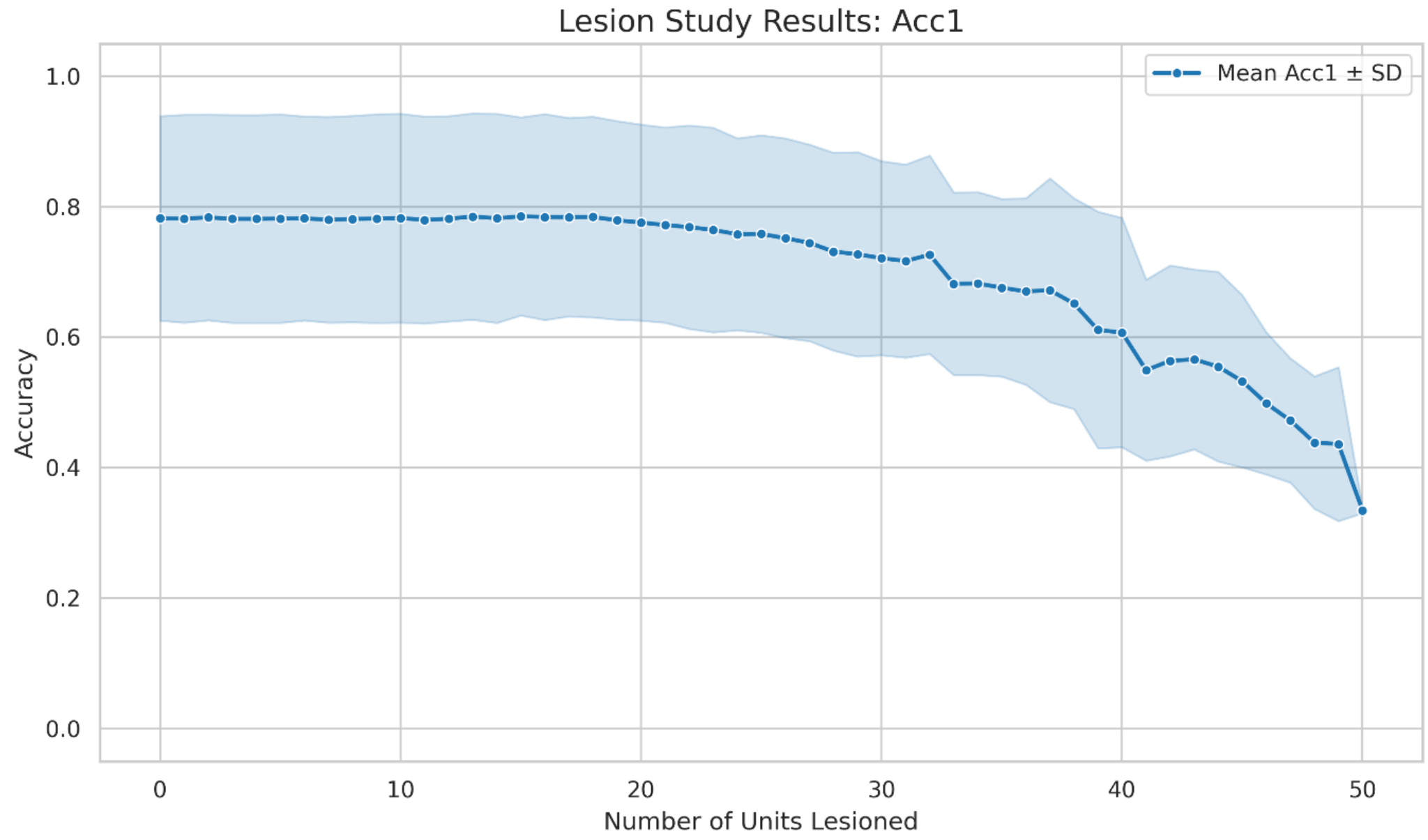

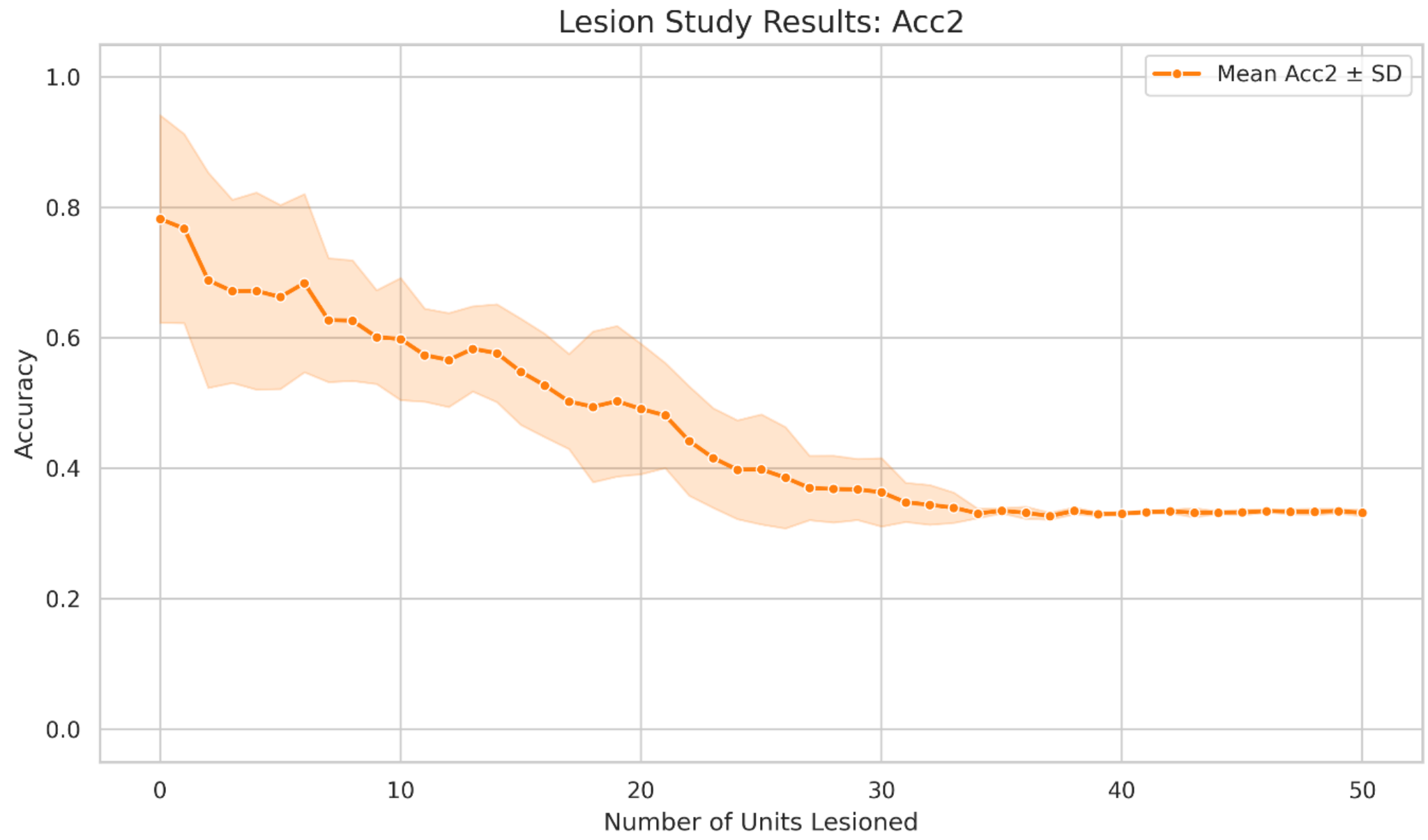

**Supplementary Fig. 1 | Cumulative lesion effects on temporal discrimination.** State-space model (SSM) accuracy was evaluated over 25,000 random trials at each lesion level. The blue curve shows lesions in ascending unit-index order (1 to 50); the orange curve shows lesions in descending order (50 to 1). Curves show the mean across four independently trained models, and shaded bands indicate ±1 standard deviation (n = 4 models).

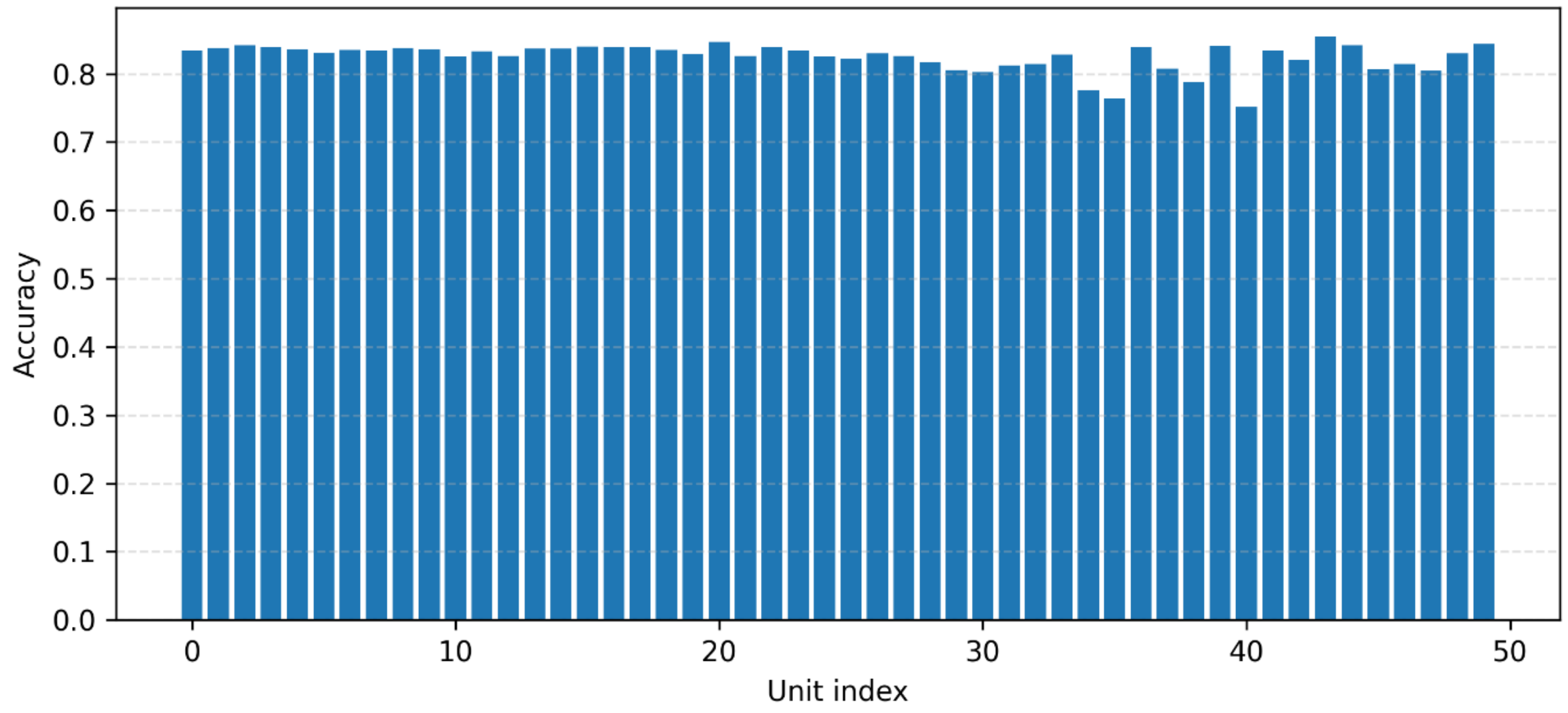

**Supplementary Fig. 2 | Effects of single-unit lesions on temporal discrimination.** Each bar shows accuracy over 25,000 random trials after lesioning the indicated hidden state unit in the state-space model used in the main text.

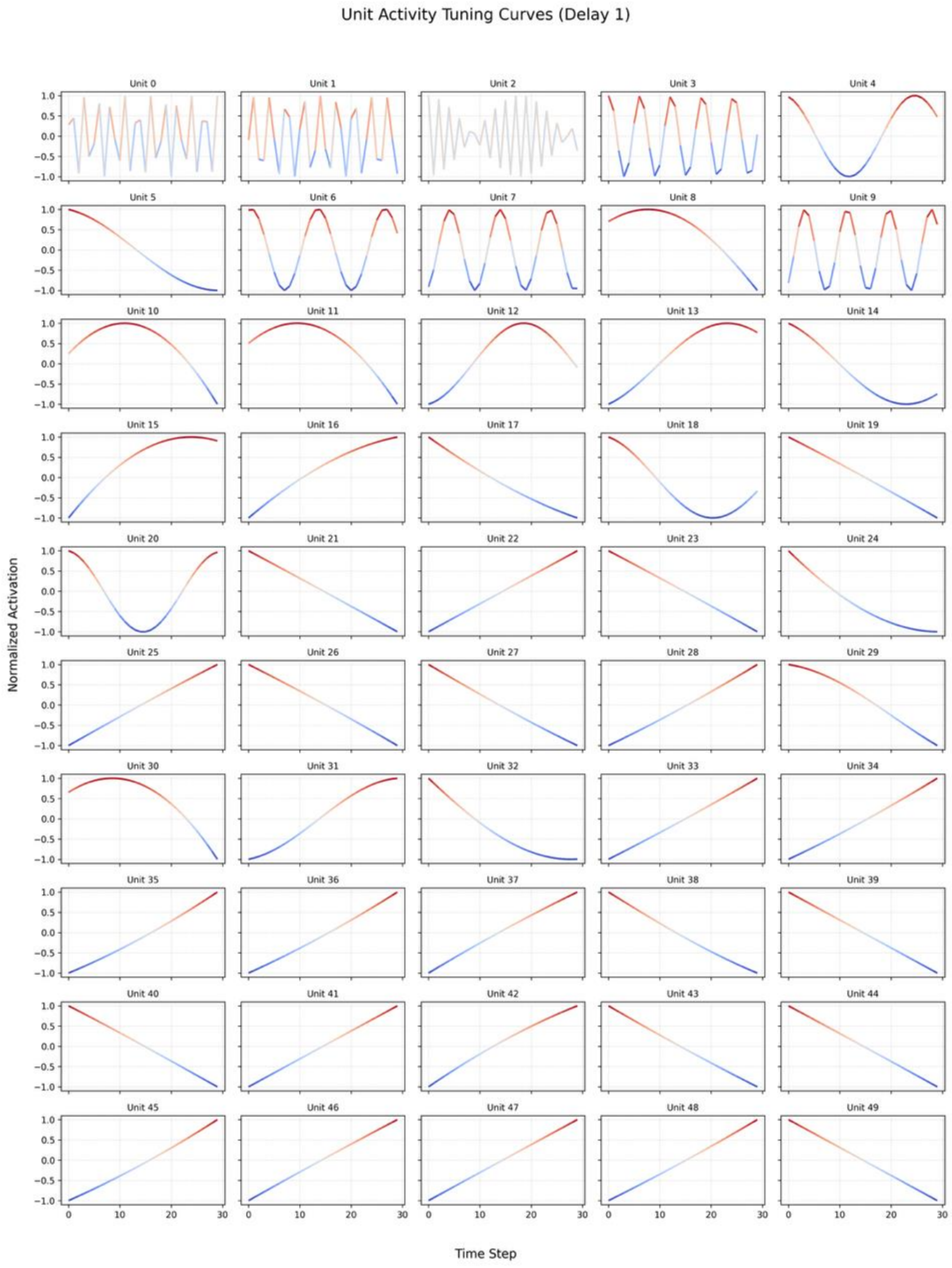

**Supplementary Fig. 3a | Individual hidden state tuning curves during the first delay period.** Units 0–49 are shown in their original index order. The second delay period is shown in Supplementary Fig. 3b on the next page.

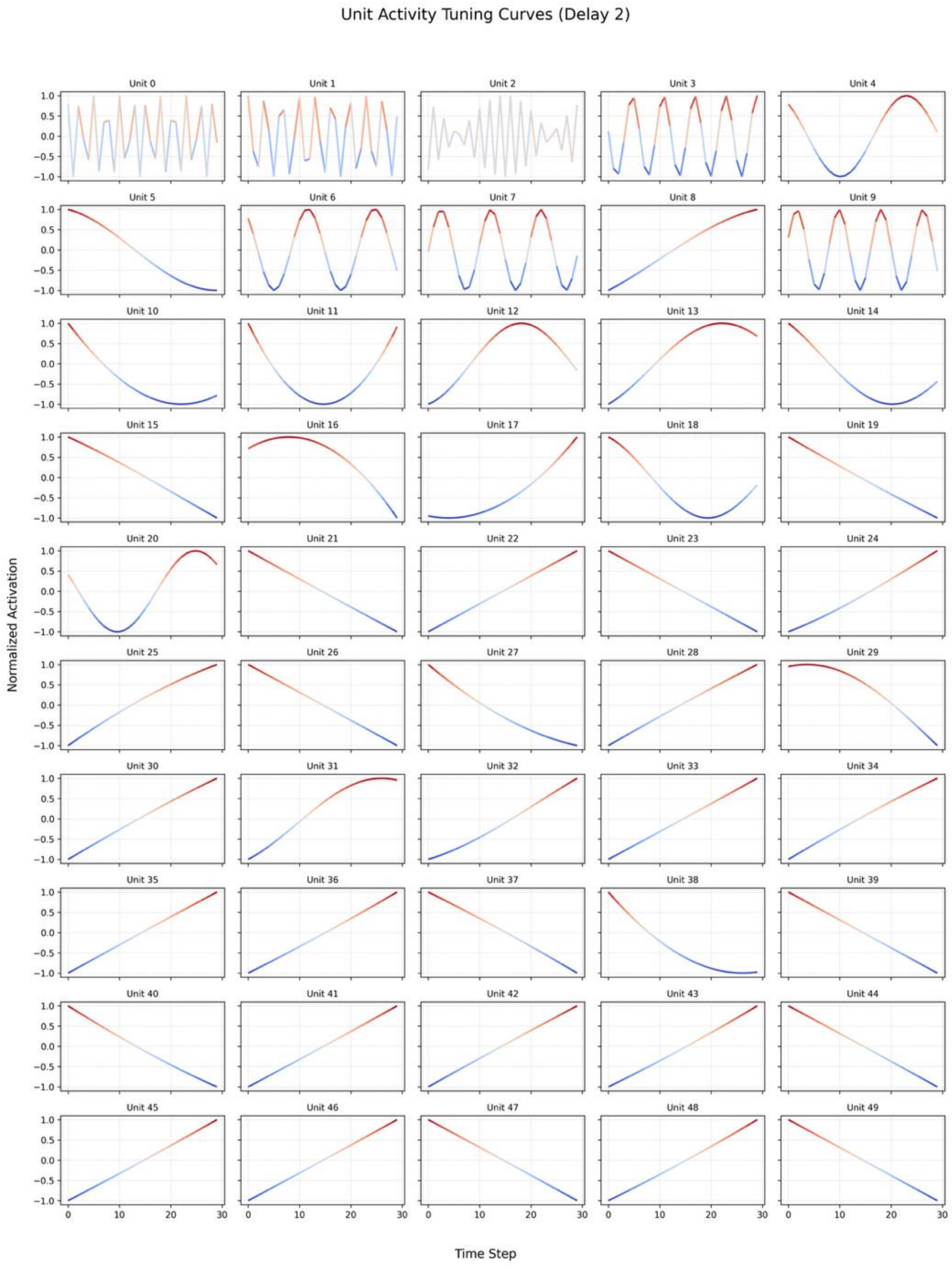

**Supplementary Fig. 3b | Individual hidden state tuning curves during the second delay period.** Units 0–49 are shown in their original index order. Each plot shows normalized, trial-averaged hidden state activation. Supplementary Fig. 3a shows the corresponding curves during the first delay period.

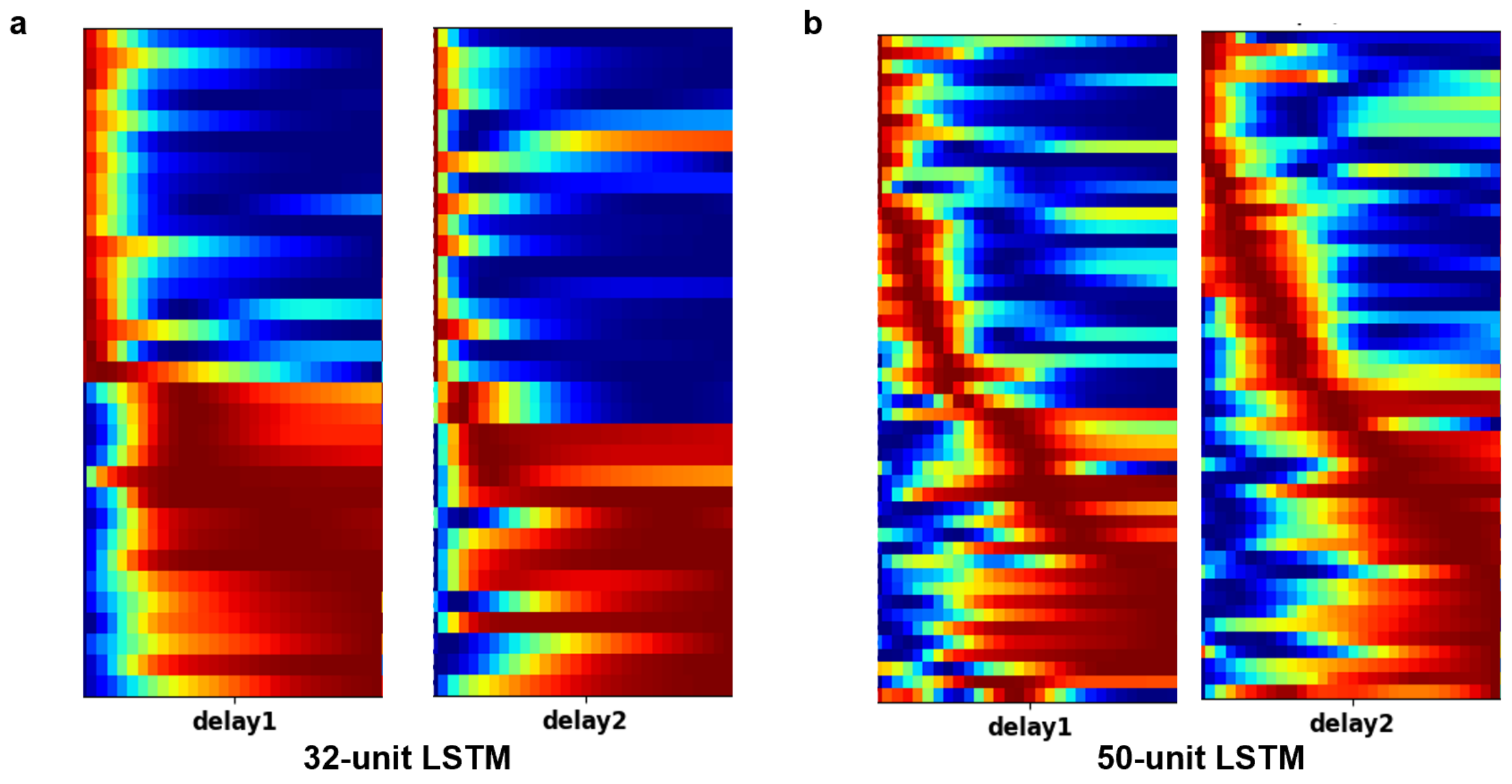

**Supplementary Fig. 4 | Hidden state dynamics of LSTM agents.** Heatmaps show trial-averaged, normalized hidden state activations during the two delay periods of the temporal discrimination task. Panels a and b show long short-term memory (LSTM) networks with 32 and 50 hidden units, respectively. Warmer colours indicate higher normalized activation and cooler colours indicate lower activation. The emerging temporal cascades accompany the performance comparison in Supplementary Table 1.

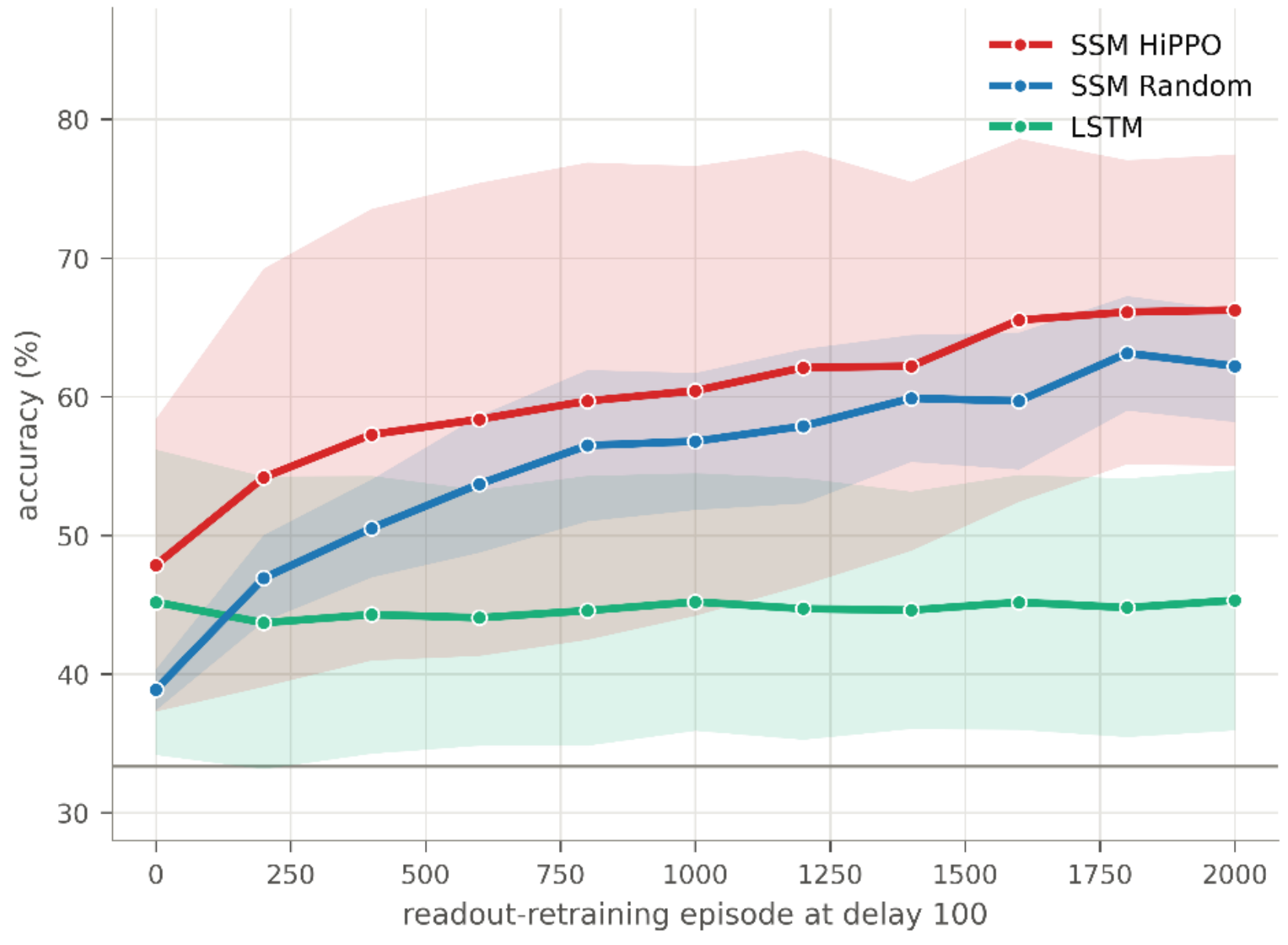

**Supplementary Fig. 5 | Recovery after extending the delay from 30 to 100 steps.** Accuracy is shown during readout retraining with the internal dynamics frozen. Red, blue and green curves show the HiPPO-initialized state-space model (SSM), randomly initialized SSM and long short-term memory (LSTM) model, respectively. Shaded bands indicate ±1 standard deviation. HiPPO denotes high-order polynomial projection operators.

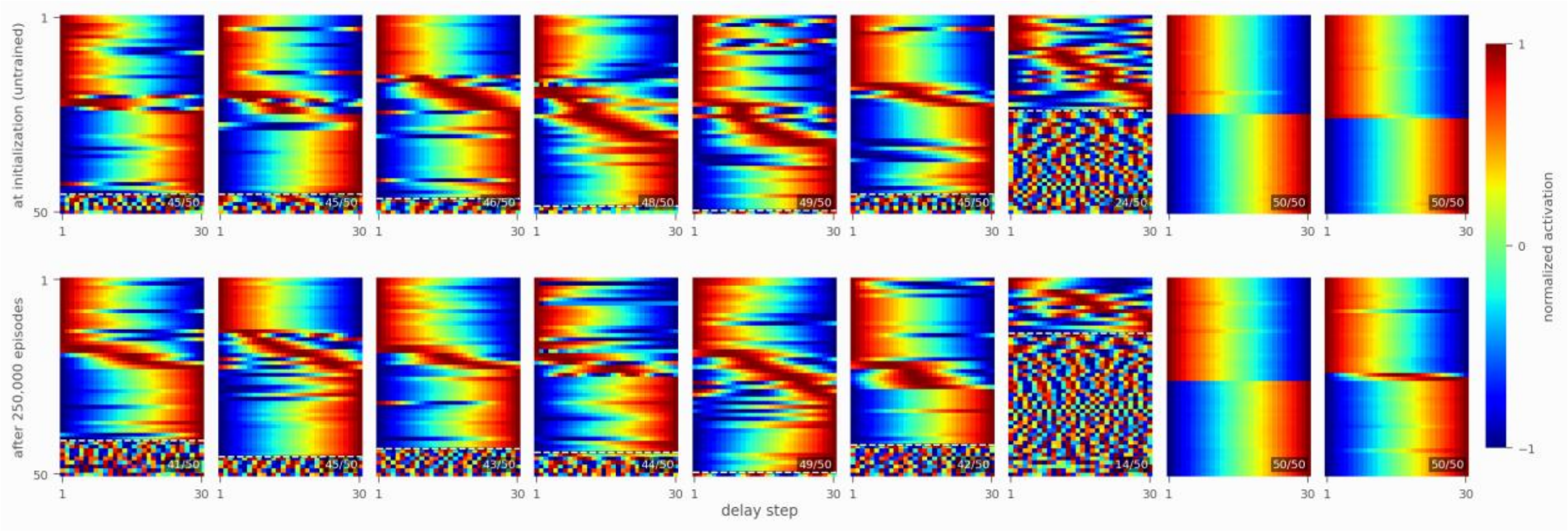

**Supplementary Fig. 6 | Hidden state dynamics across initialization schemes.** Columns show, in order, HiPPO, spectrum_matched, freq_matched, perturbed_hippo, s4d_lin, s4d_inv, alt_basis, real_diagonal and rand_complex. The first row shows the untrained models and the second row shows the trained models. Colours indicate hidden state activation, as shown by the colour scale. HiPPO denotes high-order polynomial projection operators. The initialization schemes are defined in Supplementary Note 4, and their accuracy and temporal coverage are reported in Supplementary Table 2.

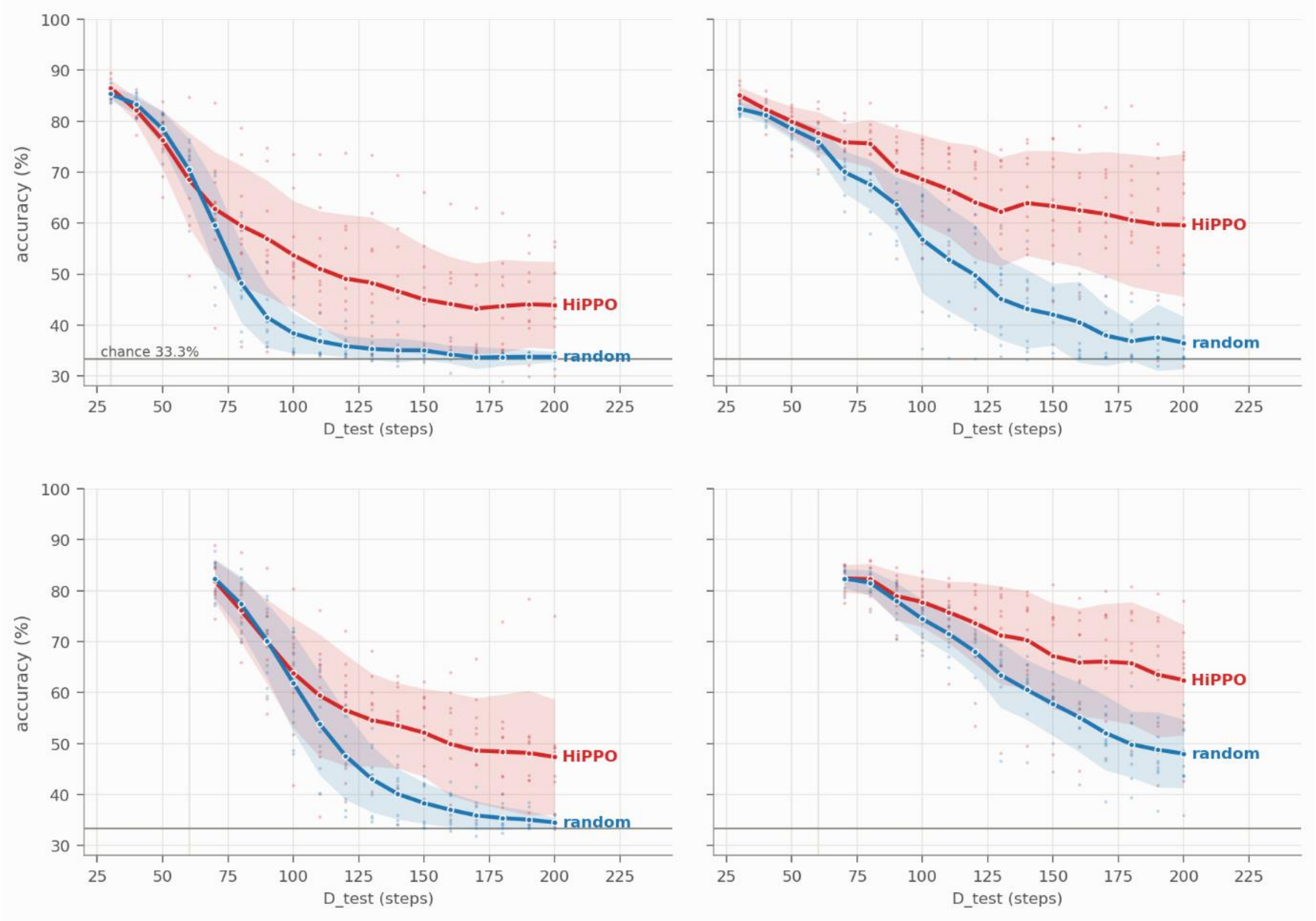

**Supplementary Fig. 7 | Retiming over extended delay durations.** The left column shows accuracy during zero-shot transfer; the right column shows recovery after retraining the readout while keeping the state-transition and input-projection parameters frozen. Red curves denote HiPPO-initialized state-space models and blue curves denote randomly initialized models. Shaded bands indicate ±1 standard deviation. The horizontal reference indicates chance performance (33.3%). The horizontal axis gives the test delay in steps. HiPPO denotes high-order polynomial projection operators.

## Supplementary Tables

**Supplementary Table 1 | Performance comparison between LSTM and SSM agents**

| Backbone | Average Performance | Note |
|---|---|---|
| LSTM H=32 | 76.3% | # Parameters matched |
| LSTM H=50 | 78.2% | # Hidden Dimension Matched |
| SSM HiPPO | 87.9% | - |
| SSM Random | 86.6% | - |

**Supplementary Table 2 | Accuracy and temporal coverage across initialization schemes**

| Initialization | | Base Accuracy @30-delay | (bins/timesteps) at initialization | (bins/timesteps) after training |
|---|---|---|---|---|
| HiPPO-based | hippo | 88.1 +/- 1.9 | 0.457 (13.7/30) | 0.452 (13.5/30) |
| | spectrum_matched | 88.9 +/- 2.2 | 0.455 (13.7/30) | 0.427 (12.8/30) |
| | freq_matched | 88.0 +/- 2.1 | 0.453 (13.6/30) | 0.422 (12.7/30) |
| | perturbed_hippo | 89.6 +/- 2.5 | 0.505 (15.2/30) | 0.450 (13.5/30) |
| | s4d_lin | 85.8 +/- 3.2 | 0.485 (14.6/30) | 0.547 (16.4/30) |
| | s4d_inv | 88.7 +/- 2.5 | 0.452 (13.5/30) | 0.437 (13.1/30) |
| | alt_basis | 65.6 +/- 12.5 | 0.702 (21.0/30) | 0.283 ( 8.5/30) |
| Random | real_diagonal | 86.8 +/- 2.1 | 0.033 ( 1.0/30) | 0.082 ( 2.5/30) |
| | rand_complex | 86.9 +/- 1.4 | 0.070 ( 2.1/30) | 0.087 ( 2.6/30) |

**Supplementary Table 3 | Architecture and training hyperparameters**

| Parameter Name | Value | Note |
|---|---|---|
| Learning Rate | 3e-4 | Weight update speed |
| Weight Decay | 1e-6 | Regularization |
| Training Episodes | 250000 | Total trials |
| Batch size | 1 | Only 1 trial at a time |
| Optimizer | Adam | Default PyTorch Adam |
| **SSM Hyperparameters** | | |
| Neurons | 50~80 | Depending on the task |
| Discretization | Zero-Order Hold (ZOH) | Default S5 discretization |
| Dropout Rate | 0.1 | Regularization |
| Initialization | HiPPO | S5 Initialization |
| Surrogate Function | Sigmoid | Most smooth surrogation |
| **RL Hyperparameters** | | |
| Policy Agent Neuron | 1 | Makes prediction/action |
| Critic Network Neuron | 1 | Estimates state reward |
| Reward Discount Factor | 0.99 | Dampens reward expectation |
| Initial Entropy | 0.3 | Linearly decays to 0 over trial |

## Supplementary References